\documentclass[10pt,aps,twocolumn,prd,showpacs,nofootinbib,noshowkeys,floatfix,preprintnumbers]{revtex4-1}

\usepackage{graphics,graphicx}
\usepackage{amsmath, amssymb}
\usepackage{multirow}
\usepackage{color}
\usepackage{hyperref}
\usepackage[normalem]{ulem}  % \sout{old text} for strikeout

\usepackage{ulem}
\usepackage{color}
\usepackage{cancel}
\usepackage{mathtools}
\usepackage{epstopdf}

\renewcommand\sout{\bgroup \color{blue} \ULdepth=-.5ex \ULset}
\def\slashchar#1{\setbox0=\hbox{$#1$}  % set a box for #1
   \dimen0=\wd0     % and get its size
   \setbox1=\hbox{/} \dimen1=\wd1  % get size of /
   \ifdim\dimen0>\dimen1   % #1 is bigger
      \rlap{\hbox to \dimen0{\hfil/\hfil}} % so center / in box
      #1     % and print #1
   \else     % / is bigger
      \rlap{\hbox to \dimen1{\hfil$#1$\hfil}} % so center #1
      /      % and print /
   \fi}

\newcommand{\dd}{\mathrm{d}}

\begin{document}

\title{Net-baryon number fluctuations in the Hybrid Quark-Meson-Nucleon model at finite density}

\author{Micha\l{} Marczenko}
\email{michal.marczenko@ift.uni.wroc.pl}
%\affiliation{Institute of Theoretical Physics, University of Wroc\l aw, Plac Maksa Borna 9, PL-50204 Wroc\l aw, Poland}
\author{Chihiro Sasaki}
\affiliation{Institute of Theoretical Physics, University of Wroc\l aw, Plac Maksa Borna 9, PL-50204 Wroc\l aw, Poland}
\date{\today}
%\affiliation{Institute of Theoretical Physics, University of Wroc\l aw,
%PL-50204 Wroc\l aw, Poland}

\begin{abstract}

	We study the mean-field thermodynamics and the characteristics of the net-baryon number fluctuations at the phase boundaries for the chiral and deconfinement transitions in the Hybrid Quark-Meson-Nucleon model. The chiral dynamics is described in the linear sigma model, whereas the quark confinement is manipulated by a medium-dependent modification of the particle distribution functions, where an additional scalar field is introduced. At low temperature and finite baryon density, the model predicts a first-, second-order chiral phase transition, or a crossover, depending on the expectation value of the scalar field, and a first-order deconfinement phase transition. We focus on the influence of the confinement over higher-order cumulants of the net-baryon number density. We find that the cumulants show a substantial enhancement around the chiral phase transition, they are not as sensitive to the deconfinement transition.

\end{abstract}

\keywords{}

\pacs{12.39.Fe,12.38.Mh,25.75.Nq}

\maketitle

%%%%%%%%%%%%%%%%%%%%%%%%%%%%%%%%%%%%%%%%%%%%%%%%%%%%%%%%%%%%%%%
\section{Introduction}
\label{sec:introduction}
%%%%%%%%%%%%%%%%%%%%%%%%%%%%%%%%%%%%%%%%%%%%%%%%%%%%%%%%%%%%%%%

	Exploration of the phase diagram of quantum chromodynamics (QCD), the theory of the strong interaction, at finite temperature and density is one of the most challenging topics in high-energy particle and nuclear physics. At vanishing density, the first-principles calculations, i.e., lattice QCD (LQCD), provide a reliable description of equation of state and a profound insight into the phase structure of QCD~\mbox{\cite{Borsanyi:2011sw, Bellwied:2015lba, Bazavov:2014pvz, Bazavov:2012jq}}.
	The recent LQCD findings~\cite{Aarts:2015mma, Aarts:2017rrl} exhibit a clear manifestation of the parity doubling structure for the low-lying baryons around the chiral crossover. The masses of the positive-parity groundstates are found to be rather insensitive to temperature, while the masses of negative-parity states drop substantially with increasing temperature, and the parity doublers become almost degenerate with a finite mass in the vicinity of the chiral crossover $T_c$. Despite unphysically heavy pion mass used in the study, this is likely an imprint of the chiral symmetry restoration in the baryonic sector of QCD, and is expected to occur also in cold dense matter. This can be described in a schematic framework with chiral symmetry, the parity doublet model~\cite{Detar:1988kn,Jido:1999hd,Jido:2001nt}. The model has been applied to hot and dense hadronic matter, neutron stars, as well as the vacuum phenomenology of QCD~\cite{Zschiesche:2006zj, Dexheimer:2007tn, Gallas:2009qp, Sasaki:2010bp, Sasaki:2011ff, Gallas:2011qp, Steinheimer:2011ea, Paeng:2011hy, Benic:2015pia, Giacosa:2011qd, Motohiro:2015taa, Weyrich:2015hha}.

	The mechanism of quark confinement and its relation to the chiral symmetry breaking are of major importance in probing the QCD phase transition, although it is non-trivial to embed their interplay into a single effective theory. A major approach is to introduce the temporal gauge field as a static external field to a chiral Lagrangian, so that the Polyakov loop naturally appears in the thermodynamics, e.g., the Polyakov loop-extended Nambu--Jona-Lasinio (PNJL)~\cite{Fukushima:2003fw, Ratti:2005jh, Sasaki:2006ww, Roessner:2006xn, Fukushima:2008wg} or Polyakov loop-extended quark-meson (PQM)~\cite{Schaefer:2007pw, Skokov1, Skokov2, Mizher:2010zb, Herbst:2010rf} models. While at finite temperature and low density the Polyakov loop expectation value serves as an approximated order parameter for the deconfinement, it is highly questionable that it remains so at high density.

	In this work we put special emphasis on low-temperature and finite-density region of QCD phase diagram. Hence, instead of using the Polyakov loop, we shall employ the hybrid quark-meson-nucleon (Hybrid QMN) model~\cite{Benic:2015pia}. The model includes quark degrees of freedom on top of hadrons, but prevents the quarks from their unphysical onset at low density. This can be achieved by a new auxiliary scalar field to which the fermions are coupled non-trivially. This field serves as a momentum cutoff in the Fermi-Dirac distribution functions, and plays a role of suppressing the unphysical thermal fluctuations of fermions, depending on density.

	%The model includes the quark degrees of freedom in the linear sigma model. The statistical confinement mechanism is realized by a simple restriction of the Fermi-Dirac distribution functions for nucleons and quarks. This is achieved in a simple model, where an additional scalar field is introduced.

	%This field serves as a momentum cutoff for the quarks at low densities and nucleons at high densities.

	%The quark momenta are suppressed below $b$, and likewise, the nucleon momenta restricted above $\alpha b$. The $\alpha$ is the parameter of the model. The finite value of $b$ is guaranteed by a vacuum potential $V_b$, which non-monotonic behavior is associated with the deconfinement transition.

	We will study the behavior of the second- and higher-order cumulants of the net-baryon number density up to the fourth order, as well as the bulk equation of state, in the Hybrid QMN model. It is systematically examined that to what extent the thermal behaviors are dominated by the chiral criticality and the onset of deconfinement.

	This paper is organized as follows. In Sec.~\ref{sec:qcd_matter}, we introduce the parity doublet model at finite density and finite temperature, as well as its extension -- the Hybrid Quark-Meson-Nucleon model -- that includes the mechanism to mimic quark confinement. We discuss the obtained numerical results on the equation of state in Sec.~\ref{sec:results}, and the results for the second and higher-order cumulants in Sec.~\ref{sec:cumulants}. Finally, Sec.~\ref{sec:conclusions} is devoted to the summary and conclusions.

%%%%%%%%%%%%%%%%%%%%%%%%%%%%%%%%%%%%%%%%%%%%%%%%%%%%%%%%%%%%%%%
\section{Formalism}
\label{sec:qcd_matter}
%%%%%%%%%%%%%%%%%%%%%%%%%%%%%%%%%%%%%%%%%%%%%%%%%%%%%%%%%%%%%%%

	In this section, we give a brief description of the Hybrid Quark-Meson-Nucleon (Hybrid QMN) model for the QCD transitions at finite temperature and density. Following Ref.~\cite{Benic:2015pia}, we first introduce the pure hadronic model of parity doublers, and then extend it incorporating the quark degrees of freedom. Throughout this paper we consider a system with $N_f = 2$.

%%%%%%%%%%%%%%%%%%%%%%%%%%%%%%%%%%%%%%%%%%%%%%%%%%%%%%%%%%%%%%%
	\subsection{Parity Doublet model with dilaton}
	\label{sec:parity_doublet}
%%%%%%%%%%%%%%%%%%%%%%%%%%%%%%%%%%%%%%%%%%%%%%%%%%%%%%%%%%%%%%%

		In the conventional Gell-Mann--Levy model of mesons and nucleons~\cite{GML}, the nucleon mass is entirely generated by the non-vanishing expectation value of the sigma field. Thus, the nucleon inevitably becomes massless when the chiral symmetry gets restored. This is led by the particular chirality-assignment to the nucleon parity doublers, where the nucleons are assumed to be transformed in the same way as the quarks are under chiral rotations.

		More general allocation of the left- and right-handed chiralities to the nucleons, the mirror assignment, was proposed in~\cite{Detar:1988kn}. This allows an explicit mass-term for the nucleons, and consequently the nucleons stay massive at the chiral restoration point. For more details, see Refs.~\cite{Detar:1988kn,Jido:1999hd,Jido:2001nt}.

		In the mirror assignment, under \mbox{$SU(2)_L \times SU(2)_R$} rotation, two chiral fields $\psi_1$ and $\psi_2$ are transformed as follows:
		\begin{equation}\label{eq:mirror_assignment}
		\begin{split}
			\psi_{1L} \rightarrow L\psi_{1L}, \;\;\;\; \psi_{1R} \rightarrow R\psi_{1R}\textrm, \\
			\psi_{2L} \rightarrow R\psi_{2L}, \;\;\;\; \psi_{2R} \rightarrow L\psi_{2R}\textrm,
		\end{split}
		\end{equation}
		where $\psi_i = \psi_{iL} + \psi_{iR}$, $L \in SU(2)_L$ and $R \in SU(2)_R$. The nucleon part of the Lagrangian in the mirror model reads
		\begin{equation}\label{eq:doublet_lagrangian}
		\begin{split}
			\mathcal{L}_N &= i\bar\psi_1\slashchar\partial\psi_1 + i\bar\psi_2\slashchar\partial\psi_2 + m_0\left(  \bar\psi_1\gamma_5\psi_2 - \bar\psi_2\gamma_5\psi_1 \right) \\
			&+ g_1\bar\psi_1 \left( \sigma + i\gamma_5 \boldsymbol\tau \cdot \boldsymbol\pi \right)\psi_1 + g_2\bar\psi_2 \left( \sigma - i\gamma_5 \boldsymbol\tau \cdot \boldsymbol\pi \right)\psi_2 \\
			&-g_\omega\bar\psi_1\slashchar\omega\psi_1 - g_\omega\bar\psi_2\slashchar\omega\psi_2 \textrm,
		\end{split}
		\end{equation}
		where $g_1$, $g_2$, and $g_\omega$ are the baryon-to-meson coupling constants and $m_0$ is a mass parameter. Since the presence of the mass spoils scale invariance, we further impose that the theory at classical level is invariant under scale transformation. To this end, we introduce a dilaton $\chi$~\cite{Schechter} and replace the above mass with
		\begin{equation}
		m_0 \, \rightarrow\, g_\chi \chi\,,
		\end{equation}
		with a new coupling $g_\chi$. The nucleon mass emerges when the dilaton field gets condensed.
		%Note that the parity doublet model accomodates for the global symmetries of QCD, i.e., the chiral symmetry and the scale invariance.

		The mesonic part of the Lagrangian reads
		\begin{equation}
		\begin{split}
			\mathcal{L}_M &= \frac{1}{2} \left( \partial_\mu \sigma\right)^2 + \frac{1}{2} \left(\partial_\mu \boldsymbol\pi \right)^2 - \frac{1}{4} \left( \omega_{\mu\nu}\right)^2  + \frac{1}{2} \left(\partial_\mu \chi\right)^2\\
							&-V_\sigma - V_\omega - V_\chi \textrm,
		\end{split}
		\end{equation}
		where $\omega_{\mu\nu} = \partial_\mu\omega_\nu - \partial_\nu\omega_\mu$ is the field-strength tensor of the vector field, and the potentials read
		\begin{subequations}\label{eq:potentials_parity_doublet}
		\begin{align}
			V_\sigma &=  \frac{\lambda}{4} \left( \sigma^2 + \boldsymbol\pi^2 - \frac{\lambda_\chi}{\lambda}\chi^2 \right)^2 - \epsilon\sigma\chi^2 \textrm,\label{eq:sigma_potential}\\
			V_\omega &=  -\frac{1}{2}m_\omega^2 \left(\frac{\chi}{\chi_0} \right)^2\omega_\mu\omega^\mu\textrm,\\
			V_\chi &=  \frac{B}{4}\left(\frac{\chi}{\chi_0}\right)^4 \left[\ln\left(\frac{\chi}{\chi_0}\right)^4 - 1 \right] \textrm. \label{eq:dilaton_potential}
		\end{align}
		\end{subequations}
		The full Lagrangian of the parity doublet model is then $\mathcal L = \mathcal L_N + \mathcal L_M$. Note that the chiral symmetry and the scale invariance are explicitly broken by the linear term in $\sigma$ (Eq.~(\ref{eq:sigma_potential})).
		% and logarythmic term in $\chi$ (Eq.~(\ref{eq:dilaton_potential})), respectively.

		The parameters $\lambda$, $\lambda_\chi$ and $\epsilon$ can be connected to the vacuum meson masses and the pion decay constant as
		\begin{equation}\label{eq:parity_params}
			\lambda_\chi = \frac{ m_\sigma^2 - 3 m_\pi^2}{2\chi_0^2} \textrm{, }\;\;\;\;  \lambda = \frac{ m_\sigma^2 - m_\pi^2 }{2f_\pi^2}\textrm{, }\;\;\;\; \epsilon = \frac{m_\pi^2 f_\pi}{\chi_0^2} \textrm,
		\end{equation}
		where the pion mass $m_\pi = 138$~MeV, the pion decay constant $f_\pi = 93$~MeV and $\chi_0$ is the vacuum expectation value of the dilaton field.
		%Since the quark content of the sigma meson is not precisely known,
		We shall treat the mass of the sigma meson, $m_\sigma$, as a parameter. The value of the constant $B$ in Eq.~(\ref{eq:dilaton_potential}) is fixed by identifying the $\chi^4$ with the gluon condensate $\langle G_{\mu\nu}G^{\mu\nu} \rangle$. Since the gluon condensate is directly proportional to the \mbox{non-vanishing} trace anomaly, it can be transmuted into $B$, assuming that the QCD vacuum energy is dominated by the dilaton potential. The constant $B$ is then estimated to be $B= ( 273-546$ MeV$)^4$, and the vacuum value of the dilaton field is obtained through the following relation~\cite{Benic:2015pia},
		\begin{equation}
			m_\chi^2 = \frac{\partial^2 V_\chi}{\partial \chi^2} \Bigg|_{\chi=\chi_0} = \frac{4B}{\chi_0^2} \textrm,
		\end{equation}
		where the lowest glueball mass $m_\chi = 1700$~MeV is taken~\cite{Sexton:1995kd,Chen:2005mg}.

		The mass eigenstates $N_\pm$ are related to the $\psi_1$ and $\psi_2$ fields as follows:
		\begin{equation}
			\begin{pmatrix}
			N_+ \\
			N_-
			\end{pmatrix}
			=
			\frac{1}{\sqrt{2\cosh \delta}}
			\begin{pmatrix}
			e^{\delta/2} & \gamma_5 e^{-\delta/2} \\
			\gamma_5 e^{\delta/2} & - e^{-\delta/2} \\
			\end{pmatrix}
			\begin{pmatrix}
			\psi_1 \\
			\psi_2
			\end{pmatrix} \textrm,
		\end{equation}
		with
		\begin{equation}
			\sinh\delta = - \frac{g_1+g_2}{2g_\chi}\frac{\sigma}{\chi} \textrm.
		\end{equation}
		In the diagonal basis, the masses of the chiral partners are given by
		\begin{equation}\label{eq:doublet_masses}
			m_\pm = \frac{1}{2} \left[ \sqrt{\left(g_1+g_2\right)^2\sigma^2+4g^2_\chi\chi^2} \mp \left(g_1 - g_2\right)\sigma \right] \textrm.
		\end{equation}
		From Eq.~(\ref{eq:doublet_masses}), it is clear that, in contrast to the naive assignment under chiral symmetry, the chiral symmetry breaking generates only the splitting between the two masses. When the symmetry is restored, the masses become degenerate according to $m_\pm(\sigma=0) = g_\chi\chi$. Note that the chirally invariant mass needs not to be a constant in the current model as the dilaton field $\chi$ is identified with the gluon condensate $\langle G_{\mu\nu}G^{\mu\nu} \rangle$ which varies with temperature and density.

		To investigate the properties of strongly-interacting dense matter, we adopt a mean-field approximation~\cite{Serot:1984ey}. Rotational invariance requires that the spatial component of the $\omega_\mu$ field vanishes, namely $\langle \boldsymbol \omega \rangle = 0$\footnote{Since $\omega_0$ is the only non-zero component in the mean-field approximation, we simply denote it by $\omega_0 \equiv\omega$.}. Parity conservation on the other hand dictates $\langle \boldsymbol \pi \rangle = 0$.
		The mean-field thermodynamic potential of the parity doublet model reads
		\begin{equation}\label{eq:thermo_potential}
			\Omega = V_\sigma + V_\omega + V_\chi + \sum_{x=\pm}\Omega_x \textrm,
		\end{equation}
		with
		\begin{equation}
			\Omega_x = \gamma_x \int\frac{\dd^3 p}{(2\pi)^3}\; T \left[ \ln\left(1 - f_x\right) + \ln\left(1 - \bar f_x\right) \right]\textrm,
		\end{equation}
		where $\gamma_\pm = 2\times 2$ denotes the spin-isospin degeneracy factor for both parity partners, and $f_x$  $(\bar f_x)$ is the particle (antiparticle) Fermi-Dirac distribution function,
		\begin{equation}\label{eq:fermi_dist_nucleon}
		\begin{split}
			f_x = \frac{1}{1+ e^{\beta\left(E_x - \mu_N\right)}} \textrm,\\
			\bar f_x = \frac{1}{1+ e^{\beta\left(E_x + \mu_N\right)}}\textrm, \\
		\end{split}
		\end{equation}
		with $\beta$ being the inverse temperature, the dispersion relation $E_x = \sqrt{\boldsymbol p^2 + m_x^2}$ and the effective chemical potential $\mu_N = \mu_B - g_\omega \omega$.

		This model is composed solely of hadronic degrees of freedom. In the next subsection we give its extension including further quark degrees of freedom.

%%%%%%%%%%%%%%%%%%%%%%%%%%%%%%%%%%%%%%%%%%%%%%%%%%%%%%%%%%%%%%%
	\subsection{Hybrid Quark-Meson-Nucleon model}
	\label{sec:HQMN_model}
%%%%%%%%%%%%%%%%%%%%%%%%%%%%%%%%%%%%%%%%%%%%%%%%%%%%%%%%%%%%%%%

		\begin{figure*}[t!]
		\begin{center}
			\includegraphics{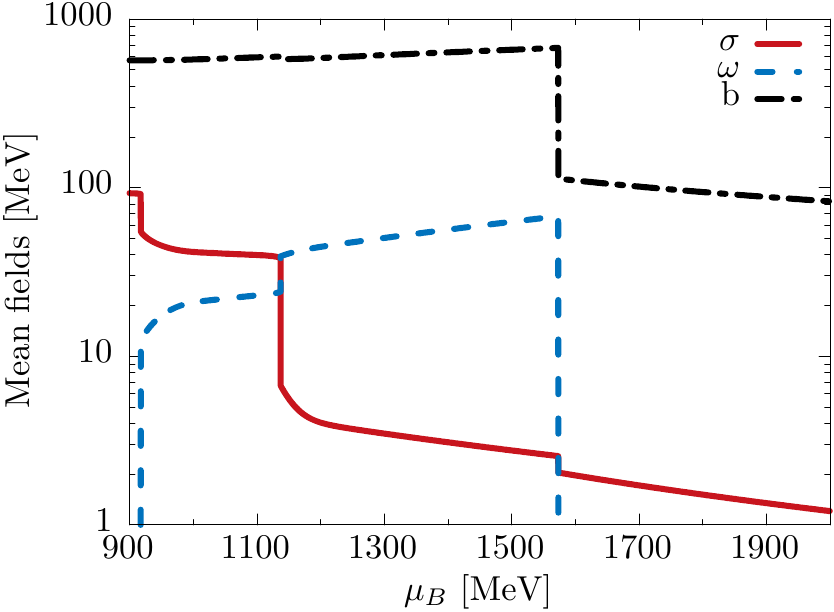}
			\includegraphics{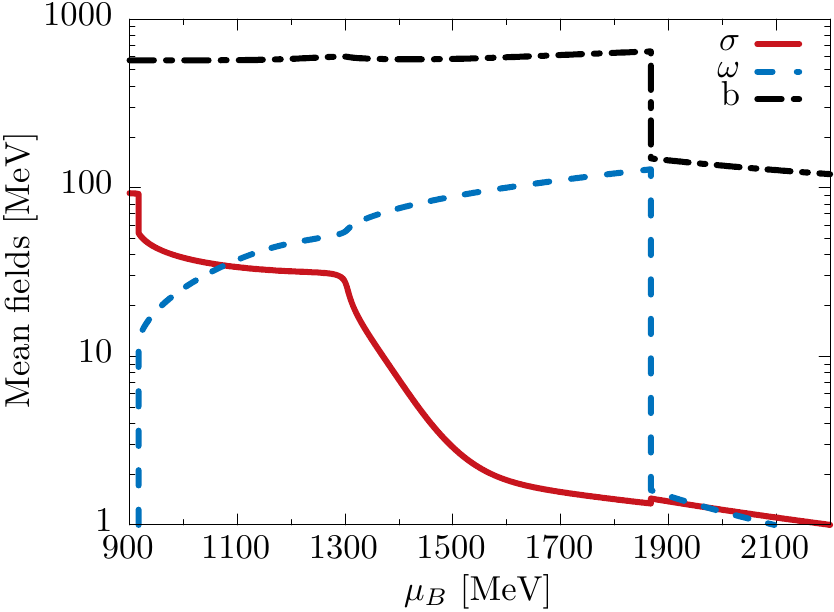}
			\caption{(Color online) The mean fields in the QMN model for $\alpha b_0 = 300$~MeV (left) and $\alpha b_0 = 390$~MeV (right) at temperature $T=10$~MeV.
			%In both, there are clear indications of the liquid-gas transition in the $\sigma$ field (red solid curve). Also, the first-order chiral transition is seen as a discontinuous change of the $\sigma$-field expectation value in the former case (left), while the crossover in the latter is seen as a continuous drop. The deconfinement transition at higher values of baryon chemical potential is assosiated with the drop in $b$ field (black dash-dotted curve) in both cases.
			}
			\label{fig:mean_fields}
		\end{center}
		\end{figure*}
		The hybrid approach proposed in~\cite{Benic:2015pia} is a natural extension of the parity doublet model in twofold meaning: First, it includes the quark degrees of freedom, and the fermionic Lagrangian in Eq.~(\ref{eq:doublet_lagrangian}) is extended with a quark-meson coupling as in the standard quark-meson model,
		\begin{equation}\label{eq:quarks_lagrangian}
		\mathcal{L}_q = i\bar q\slashchar\partial q + g_q\bar q \left( \sigma + i\gamma_5 \boldsymbol\tau\cdot\boldsymbol\pi \right) q \textrm.
		\end{equation}
		The full model Lagrangian is then
		\begin{equation}\label{eq:full_hqmn_lagrangian}
			\mathcal L = \mathcal L_N + \mathcal L_q + \mathcal L_M \textrm.
		\end{equation}
		Note that the nucleons and quarks are coupled to the same mean fields generated from the potential $V_\sigma$. On the other hand, the nature of the repulsive interaction among quarks is still far from consensus. In general, a coupling of quarks to the repulsive $\omega$ mean field can be treated as an additional parameter. In order to avoid unnecessary complexity and to reduce the number of free parameters, we do not take it into account in the current work.
		
		Second, the model realizes the concept of {\it statistical confinement}~\footnote{In a class of chiral models implementing the Polyakov loop~\mbox{\cite{MO,Fuku,MST,RTW,MAS}}, the suppression of unphysical quarks in hadronic sector is achieved by the expectation value of the Polyakov loop $\Phi$, which substantially modifies the statistical distribution function. However, the quarks do have a small contribution to the pressure even at a very low temperature. The quarks are thus unconfined at any temperature in this approach. We refer to the thermodynamic  suppression of the quarks as statistical confinement~\cite{SFR}.} through a medium-dependent modification of the Fermi-Dirac distribution functions, where an auxiliary scalar field $b$ (bag field) is introduced. The distribution functions for the nucleons are replaced with
		\begin{equation}\label{eq:fermi_dist_nucleons}
		\begin{split}
			n_\pm = \theta \left(\alpha^2b^2 - \boldsymbol p^2 \right) f_\pm \textrm,\\
			\bar n_\pm = \theta \left(\alpha^2 b^2 - \boldsymbol p^2 \right) \bar f_\pm \textrm,
		\end{split}
		\end{equation}
		and for the quarks, accordingly
		\begin{equation}\label{eq:fermi_dist_quarks}
		\begin{split}
			n_q = \theta \left(\boldsymbol p^2 - b^2 \right) f_q \textrm, \\
			\bar n_q = \theta \left(\boldsymbol p^2 - b^2 \right) \bar f_q \textrm,
		\end{split}
		\end{equation}
		where $\alpha$ is a dimensionless model parameter. The functions $f_\pm$ are given in Eq.~(\ref{eq:fermi_dist_nucleon}), and for quarks they are defined as
		\begin{equation}
		\begin{split}
			f_q = \frac{1}{1+ e^{\beta\left(E_q - \mu_q\right)}} \textrm, \\
			\bar f_q = \frac{1}{1+ e^{\beta\left(E_q + \mu_q\right)}}\textrm, \\
		\end{split}
		\end{equation}
		with the quark chemical potential $\mu_q = \frac{1}{3} \mu_B$, and the dispersion relation $E_q = \sqrt{\boldsymbol p^2 + m_q^2}$.

		The real scalar field $b$ is generated from a vacuum bag potential $V_b$. Following~\cite{Benic:2015pia}, we take the potential of the form
		\begin{equation}\label{eq:bag_potential}
			V_b = -\frac{1}{2}\kappa_b^2 b^2 + \frac{1}{4}\lambda_b b^4 \textrm.
		\end{equation}
		The potential~(\ref{eq:bag_potential}) develops a non-trivial vacuum expectation value at $b = \sqrt{\kappa_b^2 / \lambda_b}$. From Eqs.~(\ref{eq:fermi_dist_nucleons}) and (\ref{eq:fermi_dist_quarks}), one finds that the nucleons favor large $b$, whereas the quarks do small $b$. The potential~(\ref{eq:bag_potential}) is chosen such that, at a certain $T$ and $\mu_B$, a transition sets in, causing the bag-field expectation value to abruptly drop. As a consequence, at low $T$ and $\mu_B$, the quark degrees of freedom are suppressed, while the nucleons get suppressed at high $T$ and $\mu_B$. This characteristic behavior is associated with the deconfinement transition, which is a crucial feature of the model~\cite{Benic:2015pia}.

		We emphasize that the underlying symmetry of the potential $V_b$ must be discrete. This is because otherwise additional Nambu-Goldstone modes would emerge in the low-energy sector of QCD, other than pions, and they will spoil the known phenomenology of the chiral dynamics.

		%No thet, since one does not expect any new Goldstone modes in QCD, which occurence accompanies spontateuous breaking of a continuous symmetry, the potential $V_b$ ought to be constructed based on a spontaneuous breaking pattern through a discrete symmetry.

		The mean-field thermodynamic potential of the Hybrid QMN model reads
		\begin{equation}\label{eq:thermo_potential}
			\Omega = V_\sigma + V_\omega + V_\chi + V_b + \sum_{x=\pm, q}\Omega_x \textrm,
		\end{equation}
		with
		\begin{equation}
			\Omega_x = \gamma_x \int\frac{\dd^3 p}{(2\pi)^3}\; T \left[ \ln\left(1 - n_x\right) + \ln\left(1 - \bar n_x\right) \right]\textrm,
		\end{equation}
		where $\gamma_q = 2 \times N_c \times N_f$ denotes the spin-color-flavor degeneracy factor for quarks. In this study, $N_f = 2$ and $N_c = 3$, hence $\gamma_q = 12$.

		The thermal values of the mean fields are determined by minimizing the thermodynamic potential:
		\begin{subequations}\label{eq:gap_eqs}
		\begin{align}
		\frac{\partial\Omega}{\partial\sigma} &= -\lambda_\chi\chi^2\sigma + \lambda\sigma^3 - \epsilon\chi^2 + \sum_{x=\pm,q}s_x \frac{\partial m_x}{\partial \sigma} = 0 \textrm, \label{eq:gap_eq_sigma}\\
		\frac{\partial\Omega}{\partial\omega} &= -m_\omega^2 \left(\frac{\chi}{\chi_0}\right)^2\omega + g_\omega\sum_{x=\pm}\rho_x = 0 \textrm,\label{eq:gap_omega}\\
		\begin{split}
			\frac{\partial\Omega}{\partial \chi} &= -\lambda_\chi\sigma^2\chi + \frac{\lambda_\chi^2}{\lambda}\chi^3 - 2\epsilon\sigma\chi-m_\omega^2\frac{\chi}{\chi_0^2}\omega^2 \label{eq:gap_dilaton}\\
			& + B \frac{\chi^3}{\chi_0^4} \ln\left(\frac{\chi}{\chi_0}\right)^4 + \sum_{x=\pm} s_x\frac{\partial m_x}{\partial \chi} =0\textrm,
		\end{split}\\
		\frac{\partial\Omega}{\partial b} &= - \kappa_b^2 b + \lambda_b b^3 + \alpha \sum_{x=\pm} \hat\omega_x - \hat\omega_q= 0 \label{eq:gap_bag}\textrm,
		\end{align}
		\end{subequations}
		where the scalar and baryon densities read
		\begin{equation}\label{eq:scalar_den}
			s_x = \gamma_x \int\frac{\dd^3 p}{(2\pi)^3}\; \frac{m_x}{E_x} \left( n_x + \bar n_x \right) \textrm,
		\end{equation}
		and
		\begin{equation}\label{eq:scalar_den}
			\rho_x = \gamma_x \int\frac{\dd^3 p}{(2\pi)^3}\; \left( n_x - \bar n_x \right) \textrm,
		\end{equation}
		respectively. The boundary terms in the gap equation~(\ref{eq:gap_bag}) read
		\begin{equation}\label{eq:boundary_nucleon}
			\hat\omega_\pm = \gamma_\pm \frac{(\alpha b)^2}{2\pi^2} T \left[ \ln\left(1 - f_\pm\right) + \ln\left(1 - \bar f_\pm\right) \right]_{\boldsymbol p^2 = (\alpha b)^2}\textrm,
		\end{equation}
		and
		\begin{equation}\label{eq:boundary_quark}
			\hat\omega_q = \gamma_q \frac{b^2}{2\pi^2} T \left[ \ln\left(1 - f_q\right) + \ln\left(1 - \bar f_q\right) \right]_{\boldsymbol p^2 = b^2} \textrm,
		\end{equation}
		for the nucleons and quarks, respectively. Note that the terms (\ref{eq:boundary_nucleon}) and (\ref{eq:boundary_quark}) come into the gap equation (\ref{eq:gap_bag}) with opposite signs. This leads to that the nucleons and quarks favor different values of the bag field.
		\begin{figure}[t!]
		\begin{center}
			\includegraphics{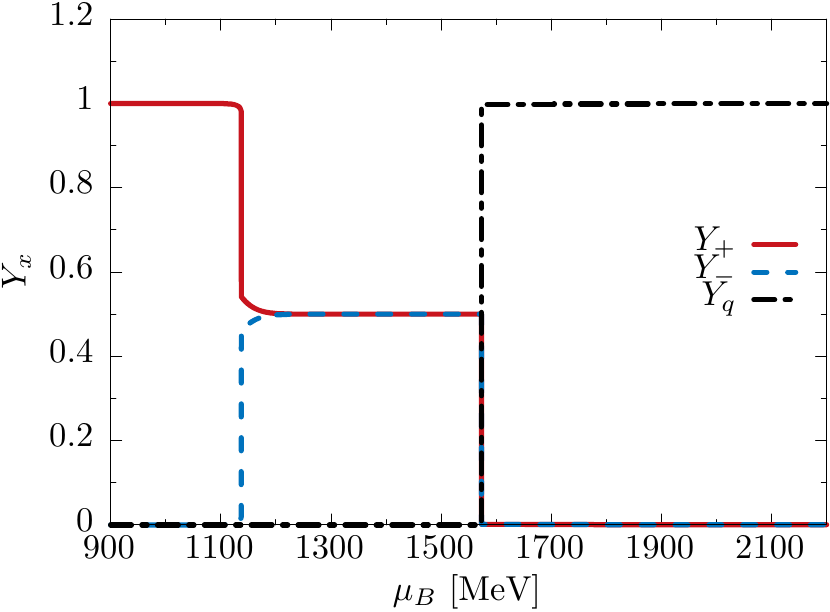}\\
			\includegraphics{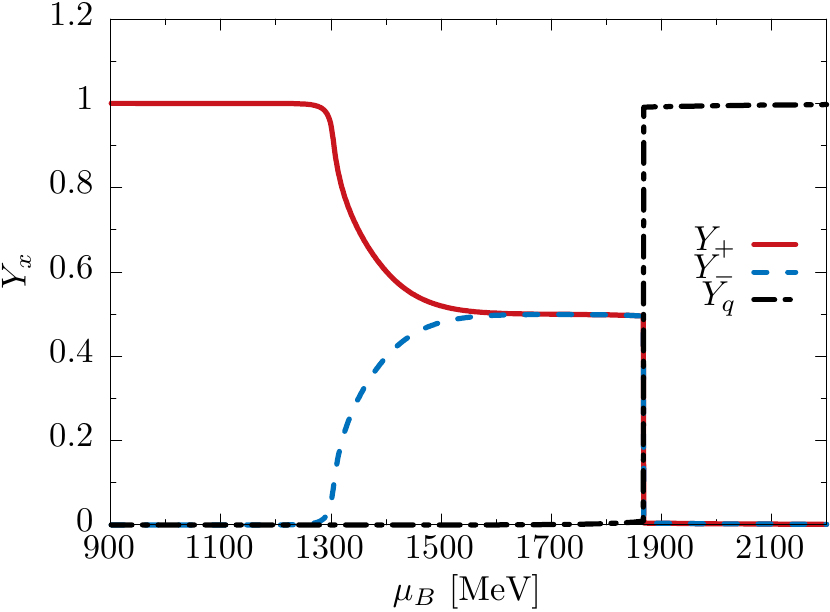}
			\caption{(Color online) Net-baryon density fractions from Eq.~(\ref{eq:particle_fraction}), plotted as functions of baryon chemical potential at $T=10$~MeV. The top panel shows the case with $\alpha b_0 = 300$~MeV.
			%In this scenario, the sudden decrease in the $N_+$ (red solid line), and corresponding enhancement of the $N_-$ (blue dashed line) density fractions, resemble the 1st-order chiral phase transition.
			The result in the bottom panel is obtained with $\alpha b_0 = 390$~MeV.
			%the chiral transition is smooth crossover, and the nucleons gradually become equally populated. The deconfinement is understood as the abrupt increase of the quark density fraction (black dash-dotted line) in both cases.
			}
			\label{fig:fractions}
		\end{center}
		\end{figure}
		
		In the grand canonical ensemble, the net-baryon number density for nucleons and quarks can be calculated as follows
		\begin{equation}\label{eq:baryon_density}
		\begin{split}
			\rho_\pm &= -\frac{\partial \Omega_\pm}{\partial \mu_B} \textrm, \\
			\rho_q &= -\frac{\partial \Omega_q}{\partial \mu_B} = -\frac{1}{3}\frac{\partial \Omega_q}{\partial \mu_q} \textrm,
		\end{split}
		\end{equation}
		respectively. The total net-baryon number density is then the sum of all the terms in Eq.~(\ref{eq:baryon_density}),
		\begin{equation}
			\rho_B = \rho_+ + \rho_- + \rho_q \textrm.
		\end{equation}
		The particle-density fractions are defined as
		\begin{equation}\label{eq:particle_fraction}
			Y_x = \frac{\rho_x}{\rho_B} \textrm.
		\end{equation}
		As will be seen in the next section, traces of both chiral and deconfinement transitions will be nicely pronounced in the above fractional net-baryon number densities.

%%%%%%%%%%%%%%%%%%%%%%%%%%%%%%%%%%%%%%%%%%%%%%%%%%%%%%%%%%%%%%%
	\subsection{Determination of model parameters}
	\label{sec:parameters}
%%%%%%%%%%%%%%%%%%%%%%%%%%%%%%%%%%%%%%%%%%%%%%%%%%%%%%%%%%%%%%%

		The original Hybrid QMN model can deal with the dilaton dynamics through a dilaton-to-fermion coupling. However, as shown in Ref.~\cite{Benic:2015pia}, due to its heavy mass, the presence of the dilaton mean-field does not affect the nuclear groundstate properties.
		%We confirm this by explicit numerical calculations.
		In fact, one finds that the expectation value of the dilaton remains nearly constant in the region of interest, and does not influence the model predictions. Hence, for the sake of simplicity of the calculations and the clarity of the discussion, we assume a constant value of the dilaton field, equal to its vacuum expectation value, $\chi_0$. As a result, the model becomes independent of the dilaton dynamics and the corresponding gap equation~(\ref{eq:gap_dilaton}) is irrelevant in the following discussion. The chirally invariant mass becomes a constant, i.e., $m_0 = g_\chi \chi_0$, and is treated as a parameter. The dilaton potential becomes irrelevant as well, thus can be omitted. Consequently, the rest of the potentials~(\ref{eq:potentials_parity_doublet}) are rewritten as follows:
		\begin{subequations}\label{eq:potentials_hybrid_no_dilaton}
		\begin{align}
			V_\sigma &= -\frac{\lambda_2}{2}\left(\sigma^2 + \boldsymbol\pi^2\right) + \frac{\lambda_4}{4}\left(\sigma^2 + \boldsymbol\pi^2\right)^2 - \epsilon'\sigma \textrm,\\
			V_\omega &= -\frac{1}{2}m_\omega^2 \omega_\mu\omega^\mu\textrm,
		\end{align}
		\end{subequations}
		where the parameters $\lambda_2$, $\lambda_4$, and $\epsilon'$ are modified to be
		%~(\ref{eq:parity_params}), namely
		\begin{equation}\label{eq:parity_params}
			\lambda_2 = \lambda_\chi \chi_0^2\textrm{, }\;\;\;\; \lambda_4 = \lambda \textrm{, }\;\;\;\; \epsilon' = \epsilon\chi_0^2 \textrm.
		\end{equation}
		%with the same vacuum parameters as in Eqs.~(\ref{eq:parity_params}).
		Note that constant terms in $V_\sigma$ were dropped. The gap equations can be derived similarly to Eqs.~(\ref{eq:gap_eqs}).

		The positive-parity state corresponds to the nucleon $N(938)$, with the vacuum mass $m_+ = 938$~MeV. Its negative-parity partner is identified as $N(1535)$, with the real part of the mass pole in the range \mbox{$1490-1530$~MeV}~\cite{Olive:2016xmw}. Here, following the previous studies in Refs.~\cite{Zschiesche:2006zj, Benic:2015pia}, we fix its mass to be $m_- = 1500$~MeV. The chirally invariant mass is determined to be $m_0 = 790$~MeV, accordingly. Such high value of $m_0$ is also supported by recent findings in full lattice QCD simulations~\cite{Aarts:2015mma, Aarts:2017rrl}, at finite temperature and vanishing baryon chemical potential, where one sees clearly that the mass of the lowest positive-parity state is largely unaffected up to the pseudo-critical temperature, while a rather strong temperature-dependence is observed in the negative-parity channel. Also, the masses of the parity doublers are found to be nearly degenerate in the vicinity of the pseudo-critical temperature. Hence, the parity-doubling structure with $m_0 \sim m_+(T=0)$ is approximately realized because of the chiral symmetry restoration.

		The parameters $g_1$ and $g_2$ are determined by the aforementioned vacuum nucleon masses and the chirally invariant mass $m_0$ through Eq.~(\ref{eq:doublet_masses}). The nuclear matter saturation properties at $T=0$ set the following two constraints
		\begin{equation} \label{eq:saturation}
		\begin{split}
			E/A\left( \mu_B = 923 \textrm{~MeV} \right) - m_+ &= -16 \textrm{~MeV,} \\
			\rho_B \left( \mu_B = 923 \textrm{~MeV} \right) &= 0.16 \textrm{~fm}^{-3}\textrm.
		\end{split}
		\end{equation}
		The above are used to fix the value of the vector coupling $g_\omega$ and the mass of sigma meson $m_\sigma$. The value of the quark coupling $g_q$ is fixed by assuming that the mass of the nucleon is \mbox{$m_+ = 3 m_q$} in the vacuum.

		For the parametrization of the bag potential $V_b$, we adopt the values suggested in Ref.~\cite{Benic:2015pia}. The $\alpha$-dependence is to be studied in the next sections. Its value is chosen in such a way that the UV cutoff $\alpha b_0$ for the nucleon distribution function and the IR cutoff $b_0$ for the quark distribution function do not spoil the properties of the nuclear groundstate. This roughly sets the lower bound for $\alpha b_0$ to be \mbox{$\gtrsim 300$}~MeV. On the other hand, even though the quark degrees of freedom are suppressed through Eq.~(\ref{eq:fermi_dist_quarks}), they can still be thermally excited at finite temperature, even before the point where the bag field effectuates the transition. Therefore, we impose that the quark density cannot exceed 1$\%$ of the total density of the system ($Y_q \leq 0.01$) before the deconfinement transition point. This prescription sets an upper limit for the $\alpha$ parameter. In general, the limit, set by such a constraint, should decrease with increasing temperature. This is due to the fact that at higher temperature quarks are more readily excited at lower values of baryon chemical potential. Hence, the point at which the constraint is met shifts towards lower values.

		The model parameters to be used in the following discussion are summarized in Table~\ref{tab:par}. In Section~\ref{sec:results}, we discuss the role of the $\alpha$ parameter and its impact on the mean-field dynamics of the Hybrid QMN model, i.e, the chiral and deconfinement transitions, as well as various thermodynamic observables, including higher-order cumulants of the net-baryon number density.

		\begin{table}[h!]
		\begin{center}
		\begin{tabular}{|c|c|c|c|c|c|c|c|c|c|c|c|c|}
			\hline
			$m_0$ [MeV] & $m_\sigma$ [MeV] & $g_1$ & $g_2$ & $g_\omega$ & $g_q$ & $\kappa_b$ [MeV] & $\lambda_b$ \\
			\hline
			\hline
			790         & 367.92       &  13.00 & 6.96  & 6.83      & 3.36  &  155 & 0.074 \\
			\hline
		\end{tabular}
		\end{center}
		\caption{The parameters of the Hybrid \mbox{Quark--Meson--Nucleon} model used in the current work.}\label{tab:par}
		\end{table}

%%%%%%%%%%%%%%%%%%%%%%%%%%%%%%%%%%%%%%%%%%%%%%%%%%%%%%%%%%%%%%%
\section{Equation of State}
\label{sec:results}
%%%%%%%%%%%%%%%%%%%%%%%%%%%%%%%%%%%%%%%%%%%%%%%%%%%%%%%%%%%%%%%

	In this section, we study the influence of the external parameter $\alpha$ on the thermodynamic quantities in the Hybrid QMN model. In Ref.~\cite{Benic:2015pia}, the thermodynamics of the model was studied at $T=0$~MeV, already indicating a sensitivity of the order of the chiral phase transition to the choice of $\alpha$ parameter. In the current work, we study this dependence at finite temperature.

	\begin{figure}[t!]
	\begin{center}
		\includegraphics{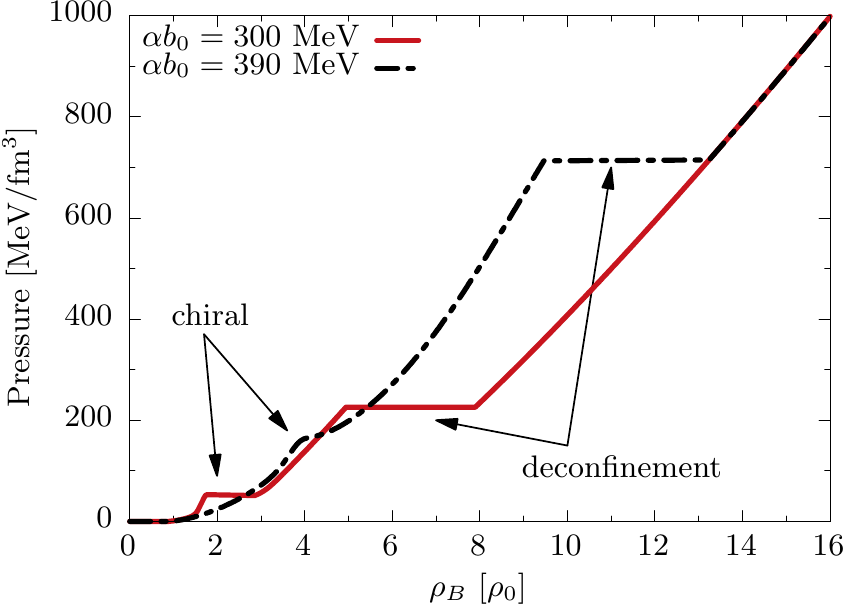}
		\caption{(Color online) Thermodynamic pressure plotted against the baryon density $\rho_B$ in the units of saturation density $\rho_0$, at $T=10$~MeV for two values of the $\alpha$ parameter. Also, the chiral and deconfinement transition points are indicated by the arrows.}
		\label{fig:pressure}
	\end{center}
	\end{figure}

	\begin{figure*}[t!]
	\begin{center}
		\includegraphics{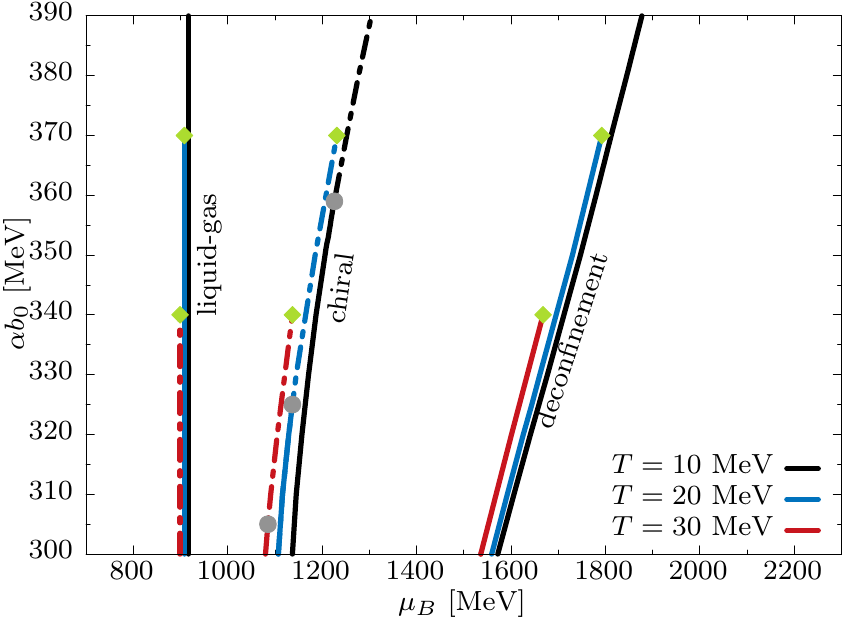}\;\;\;\;%\includegraphics[scale=.7]{phase_diagram.pdf}
		\includegraphics{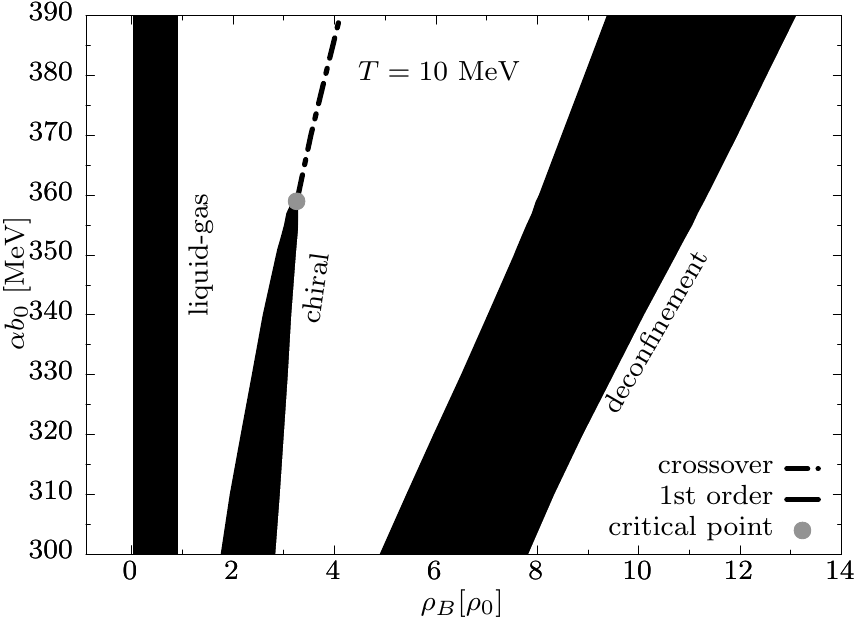}%\includegraphics[scale=.7]{phase_diagram.pdf}
		\caption{(Color online) Left: Phase diagram of the Hybrid QMN model in the \mbox{($\alpha$, $\mu_B$)-plane} at several fixed $T$. At low temperatures, the liquid-gas transition is of first order and develops the critical point (not shown here) at around $T=27$~MeV~\cite{Sasaki:2010bp}, and eventually becomes a crossover.
		%The order of the chiral phase transition is driven by the choice of the $\alpha$ parameter, and turns from the first-order transition to the crossover with increasing $\alpha$. On the other hand, the deconfinement transition is of first order by construction. For given temperature, the upper end of the transition line corresponds to the maximal values of $\alpha b_0$ allowed due to the constraint (see text), which tend to decrease with increasing temperature (see text).
		Right: Corresponding phase diagram at $T=10$~MeV in the \mbox{($\alpha$, $\rho_B$)-plane}. In the left panel, the first-order phase transitions are indicated by the solid curves. The black areas in the right panel correspond to the density jump associated with the first-order phase transition. In both panels, the broken-dashed lines indicate the crossover transition. The circles indicate the critical points on the chiral transition lines. Also, in the left panel, the diamonds indicate the points above which the constraint introduced in Sec.~\ref{sec:parameters} breaks down (see text).}
		\label{fig:alpha_diagram}
	\end{center}
	\end{figure*}

	In Fig.~\ref{fig:mean_fields}, we show the calculated mean-field expectation values at a fixed temperature, $T = 10$~MeV, for two values of the $\alpha$ parameter, namely $\alpha b_0 = 300$~MeV (left panel) and $\alpha b_0 = 390$~MeV (right panel). In both cases, the results exhibit similar behavior in the vicinity of the liquid-gas phase transition since the momentum cutoff was introduced not to spoil any properties of the nuclear ground state. In the former case, the onset of chiral restoration occurs at low baryon chemical potential $\mu_B = 1137$~MeV. After this point, the $\sigma$ expectation value suddenly drops to nearly zero, causing the parity-doublet nucleons to become almost equally populated. This is clearly seen in the top panel of Fig.~\ref{fig:fractions}, where shown are the particle abundances defined in Eq.~(\ref{eq:particle_fraction}). The chiral phase transition is, in this case, a first-order. At this point, quarks are still confined up to the point where the bag field expectation value develops a jump, which triggers the nucleon suppression, allowing, in turn, for the quarks to be populated. This happens at $\mu_B = 1573$~MeV.

	In the case of the latter scenario (right panel of Fig.~\ref{fig:mean_fields}), the chiral transition is a smooth crossover. The transition point, defined as a peak in $\partial \sigma / \partial \mu_B$, is located at \mbox{$\mu_B = 1305$~MeV}. This behavior is also resembled in the density fractions, shown in the bottom panel of Fig.~\ref{fig:fractions}. The deconfinement transition, as in the former case, is induced by the jump in the bag field expectation value and happens at a higher chemical potential, namely at $\mu_B = 1868$~MeV.

	The chiral and deconfinement transitions are reflected in the thermodynamic pressure plotted against the net-baryon number density. This is shown in Fig.~\ref{fig:pressure}. While the deconfinement in both cases is pronounced in the density jump of the order of $3-4$ $\rho_0$, the chiral transition in the case of $\alpha b_0 = 300$~MeV is seen as a slightly smaller density jump, roughly of the order of $\rho_0$. In case of $\alpha b_0 = 390$~MeV there is no true chiral transition, but a crossover, which only softens the thermodynamic pressure.
	
	In the left panel of Fig.~\ref{fig:alpha_diagram} we show different temperature profiles of the phase diagram in the \mbox{($\alpha$, $\mu_B$)-plane}. From the figure, it is clear that, due to the constraint introduced in Sec.~\ref{sec:parameters}, increasing temperature limits the range of the $\alpha$ parameter and shifts its maximal value, as well as the position of the critical point on the chiral transition line towards lower values. The full range of the parameter at $T=10$~MeV is $\alpha b_0 = 300-390$~MeV. For $T=20$~MeV, the upper limit shifts to $\alpha b_0=370$~MeV, and for $T=30$~MeV to $\alpha b_0=340$~MeV. The constraint is met with the lower bound from above at $T=37$~MeV. As seen from the figure, the phase structures at those temperatures are qualitatively similar. 
	
	In the right panel of Fig.~\ref{fig:alpha_diagram}, we show the phase diagram in the \mbox{($\alpha$, $\rho_B$)-plane}, at $T=10$~MeV. The liquid-gas transition, as argued before, is not affected by the $\alpha$ parameter. Clearly, this is so on the figure. One of the major observations is that the chiral and deconfinement transitions are always separated by about $3-5$ $\rho_0$ or $500-600$~MeV baryon chemical potential (see the left panel of Fig.~\ref{fig:alpha_diagram}). The deconfinement transition, driven by the non-dynamical bag field, is always of the first-order transition. This is due to the fact that the potential~(\ref{eq:thermo_potential}) develops a sixth-order term $b^6$ at low temperatures.

	We note that the above phase structure is modified when the chiral invariant mass $m_0$ changes. As $m_0$ is decreased, the chiral and deconfinement transitions tend to take place closer to each other and eventually become almost simultaneous with \mbox{$m_0 \approx 600$~MeV} for higher values of $\alpha$. Such transition is then necessarily of the first-order, driven by the jump in the bag-field expectation value, and triggered at lower densities.

%%%%%%%%%%%%%%%%%%%%%%%%%%%%%%%%%%%%%%%%%%%%%%%%%%%%%%%%%%%%%%%
\section{Net-baryon number density cumulants}
\label{sec:cumulants}
%%%%%%%%%%%%%%%%%%%%%%%%%%%%%%%%%%%%%%%%%%%%%%%%%%%%%%%%%%%%%%%

	The fluctuations of conserved charges reveal more information about the matter composition than the equation of state, and can be used as probes of a phase boundary. The critical properties of chiral models are governed by the same universality as in QCD, i.e., the chiral transition belongs to the $O(4)$ universality class, which, at large values of the baryon chemical potential, may develop a $Z(2)$ critical point, followed by the first-order transition~\cite{Asakawa:1989bq,Halasz:1998qr,cp}. This criticality is naturally encoded in the hadronic parity doublet model, as well as in \mbox{quark-based} models~\cite{Schaefer:2006ds,Skokov1,Skokov2,Skokov3,Almasi:2017bhq}. Recall that the Hybrid QMN model, not only includes hadron and quark degrees of freedom, but also implements the statistical confinement, through the introduction of the auxiliary bag field. It is then constructive to explore the impact of both, the chiral symmetry and the statistical confinement on the higher-order cumulants in the vicinity the two transition lines, as well as to study their asymptotic behavior.

	In the grand canonical ensemble, the cumulants are defined as higher-order derivatives of the thermodynamic potential with respect to different chemical potentials. In the current work, we are interested solely in the \mbox{net-baryon} number cumulants. They, and their rations, are defined as follows:
	\begin{equation}\label{eq:cumulants_def}
		\chi_{n} = -\frac{\partial^n \Omega \left(\mu_B, T\right)}{\partial \mu_B^n}\textrm,\;\;\;\;R_{n, m} = \frac{\chi_n}{\chi_m} \textrm.
	\end{equation}
	They are often normalized by temperature,
	%Typically, those are studied as temperature-normalized dimensionless quantities, namely
	\begin{equation}\label{eq:temperature_normalization}
		\chi_n^T = T^{n-4}\chi_n \textrm,
	\end{equation}
	as done in lattice QCD studies~\cite{lat1,lat2}.
	%They are extremely useful, e.g., in the context of heavy-ion collisions.
	%In the current context, however, more useful is the normalization by the chemical potential,
	Alternatively, one normalizes them by the chemical potential as
	\begin{equation}\label{eq:chemical_potential_normalization}
		\chi_{n}^{\mu_B} = \mu_B^{n-4}\chi_{n} \textrm,
	\end{equation}
	which are more useful in our study focusing on a cold and dense medium.
	In Appendix~\ref{sec:appendix}, we summarize the differences between those two normalizations, and discuss their restrictions and applicability in probing different regimes of the QCD phase diagram.

%%%%%%%%%%%%%%%%%%%%%%%%%%%%%%%%%%%%%%%%%%%%%%%%%%%%%%%%%%%%%%%
	\subsection{Critical behavior}
%%%%%%%%%%%%%%%%%%%%%%%%%%%%%%%%%%%%%%%%%%%%%%%%%%%%%%%%%%%%%%%

		\begin{table*}[t!]
		\begin{center}
		\begin{tabular}{|c||c|c|c|c|c||c|c|c|}
			\hline
			\multicolumn{6}{|c||}{parity doublet} & \multicolumn{3}{c|}{NJL}  \\
			\hline
			Set & $m_0$ [MeV] & $m_\sigma$ [MeV] & $g_1$ & $g_2$ & $g_\omega$ & $\Lambda$ [MeV] & $G\Lambda^2$ & $m_q$ [MeV] \\
			\hline
			\hline
			A & 790         & 367.92       &  13.00 & 6.96  & 6.83 & 900 & 1.9 & 5.6\\
			\hline
			B & 850         & 319.31       &  13.00 & 6.96  & 5.18 & 800 & 2.24 & 5.6 \\
			\hline
		\end{tabular}
		\end{center}
		\caption{The parameters for the parity doublet and NJL models. The parameters from Set~A are used for the comparison with the $\alpha b_0 = 390$~MeV scenario, and the parameters from Set~B with the $\alpha b_0 = 300$~MeV case, accordingly. In the parity doublet, Set~A is taken from~\cite{Zschiesche:2006zj}, while Set~B is chosen such that the model yields a first-order chiral phase transition at $10$~MeV temperature. Similarly to the latter case, the parametrization in the NJL model is chosen such that the model produces a chiral transition of the same order and strength comparable to the Hybrid QMN model.}\label{tab:parity_doublet_params}
		\end{table*}

		\begin{figure}[t!]
			\includegraphics{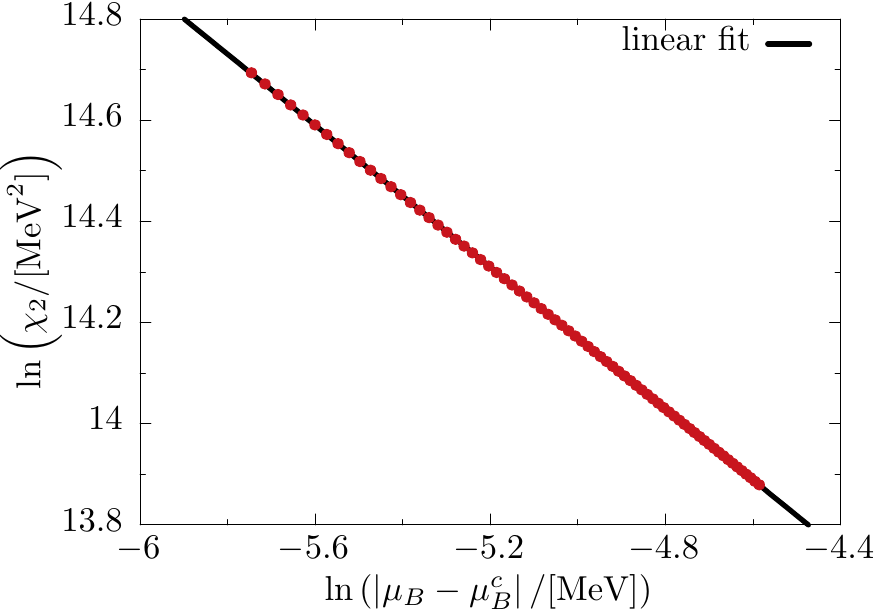}
			\caption{(Color online) The logarithm of the normalized \mbox{net-baryon} number fluctuations $\chi_2$ as a function of logarithm of $\left| \mu_B - \mu_B^c\right|$ near the CP in the mean-field approximation at $T=10$~MeV. The solid line indicates the linear fit to the values obtained numerically.}
			\label{fig:critical_exp}
		\end{figure}

		%Generalized susceptibilities of conserved charges, given as the higher-order derivatives of the thermodynamic potential with respect to different chemical potentials, reveal more information about the state of the matter.
		It is expected that, in the vicinity of phase boundaries, and especially the critical point (CP), the generalized susceptibilities of conserved charges exhibit \mbox{non-monotonic} behavior. This is governed by the singular part of the free energy and should be universal in all models that embody the QCD symmetry.

		In the following we focus on a particular example of the second-order cumulant $\chi_2$, and discuss its properties in the vicinity of phase boundaries at $T=10$~MeV and various $\mu_B$ in the mean-field approximation. At fixed temperature, the order of the chiral phase transition in the Hybrid QMN model is tuned by the model parameter $\alpha$. As seen in Fig.~\ref{fig:alpha_diagram}, for a given set of temperatures, small values of the $\alpha$ yield a first-order chiral phase transition, which eventually goes through a critical point and turns into a crossover for larger $\alpha$. In general, the position of the CP in the \mbox{($\alpha$, $\mu_B$)-plane} depends on the temperature. At the CP, irrespectively of the temperature, the susceptibility of the net-baryon number density, $\chi_2$, should diverge with the critical exponent of the 3-dimensional Ising model~\cite{SRS,Hatta:2002sj}.

		%The Parity Doublet model itself already embodies the expected $O(4)$ QCD phase structure. To confirm this behavior in the Hybrid QMN model,
		To see this in the Hybrid QMN model, we calculate the critical exponent for $\chi_2$ around the CP driven by the chiral dynamics. We take a path with fixed temperature of $T=10$~MeV and parallel to the baryon chemical potential axis, approaching the CP from smaller values. In the mean-field approximation one expects the critical exponent $\epsilon$ to be $2/3$, as the model-independent prediction.
		%This result should be qualitatively model independent~\cite{Pisarski:1983ms,Halasz:1998qr,Asakawa:1989bq,Hatta:2002sj,Schaefer:2006ds}.
		The exponent $\epsilon$ is obtained through a logarithmic fit,
		\begin{equation}\label{eq:logarythmic_fit}
		\ln{\chi_2} = -\epsilon \ln{\left|\mu_B - \mu_B^c\right|} + r \textrm,
		\end{equation}
		where $r$ is a constant.

		The value obtained numerically, $\epsilon \simeq 0.67$, stays in good agreement with the mean-field value $\epsilon = 2/3$. Fig.~\ref{fig:critical_exp} shows the numerically obtained net-baryon number susceptibility near the critical point, together with the fitted function given in Eq.~(\ref{eq:logarythmic_fit}).

%%%%%%%%%%%%%%%%%%%%%%%%%%%%%%%%%%%%%%%%%%%%%%%%%%%%%%%%%%%%%%%
	\subsection{Baryon number susceptibility}
%%%%%%%%%%%%%%%%%%%%%%%%%%%%%%%%%%%%%%%%%%%%%%%%%%%%%%%%%%%%%%%

		\begin{figure*}[t]
		\begin{center}
			\includegraphics{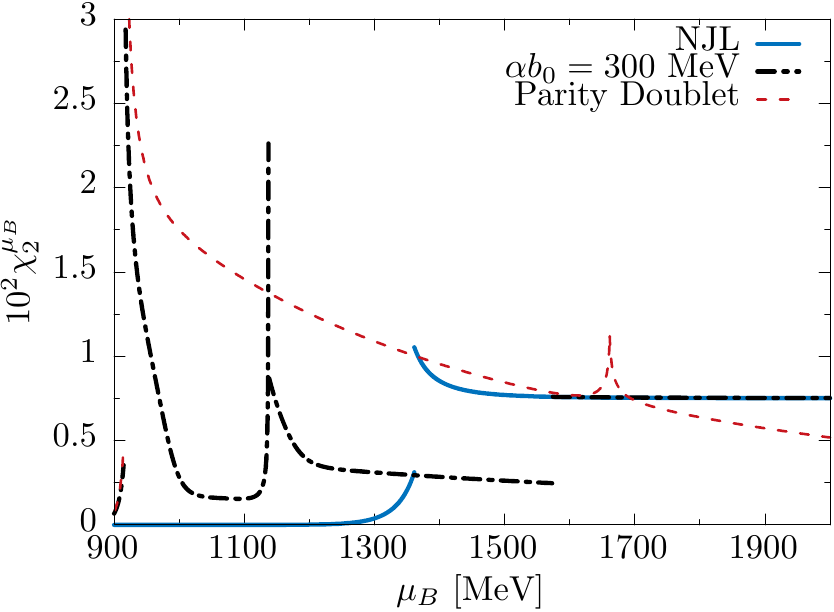}
			\includegraphics{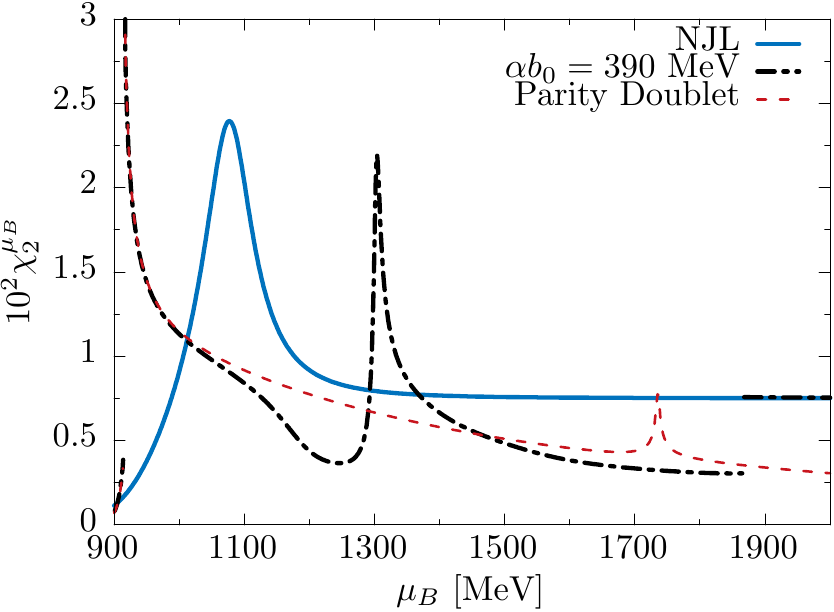}
			\caption{(Color online) Net-baryon number fluctuations normalized by $\mu_B^2$ as a function of baryon chemical potential in three models. The parameters for the parity doublet and NJL models used are tabularized in Table \ref{tab:parity_doublet_params}. The right panel corresponds to Set A, and the left panel to Set B.}
			\label{fig:x2}
		\end{center}
		\end{figure*}

		Here, we compare the results obtained in the Hybrid QMN model with those in the parity doublet model, introduced in Sec.~\ref{sec:parity_doublet}. For the sake of completeness of the discussion, we show the results from the \mbox{Nambu--Jona-Lasinio} (NJL) model in which the relevant degrees of freedom are only constituent quarks. The mean-field thermodynamic potential of the NJL model in the three-momentum cutoff scheme is given by~\cite{Klevansky:1992qe, Buballa:2003qv}
		\begin{equation}\label{eq:njl_thermodynamic_potential}
		\begin{split}
			\Omega_{\rm NJL} =& \frac{\left(M-m_q\right)^2}{4G} -\frac{N_cN_f}{\pi^2} \int\limits_0^\Lambda \dd p \; p^2 E_q\\\
			-& \frac{N_cN_f}{\pi^2}\frac{1}{\beta}\int\limits_0^\infty \dd p \; p^2 \ln\left(1+ e^{-\beta \left(E_q - \mu_q\right)}\right) \\
			-& \frac{N_cN_f}{\pi^2}\frac{1}{\beta}\int\limits_0^\infty \dd p \; p^2 \ln\left(1+ e^{-\beta \left(E_q + \mu_q\right)}\right) \textrm,\\
		\end{split}
		\end{equation}
		where $\beta$ is the inverse temperature, the bare quark mass is denoted by $m_q$, the constituent mass by $M$, \mbox{$E_q = \sqrt{p^2 + M^2}$} is the dispersion relation, and the Fermi-Dirac distributions are given by Eq.~(\ref{eq:fermi_dist_quarks}). The color and flavor degeneracy factors are $N_c = 3$ and $N_f = 2$, respectively. The constants $G$ and $\Lambda$ are external model parameters. The physical properties of the model are obtained by extremizing Eq.~(\ref{eq:njl_thermodynamic_potential}) with respect to the order parameter $M$. Finally, the higher-order cumulants are calculated according to Eq.~(\ref{eq:cumulants_def}).
		
		The parameters used in the parity doublet model and in the NJL model are tabularized in Table~\ref{tab:parity_doublet_params}. The parametrization in the NJL model is chosen so that it yields the same order of the chiral transition, with its strength comparable to the one obtained in the Hybrid QMN model on the level of $\chi_2$ observable.
		
		The characteristics of the second-order cumulants, normalized by the baryon chemical potential, $\chi_2^{\mu_B}$, in all three models are shown in Fig.~\ref{fig:x2}. In the right panel, we show the results for \mbox{$\alpha b_0 = 390$~MeV}, compared to the results obtained in other two models. The NJL result, shown as the blue solid line, undergoes a chiral crossover transition, seen as the peak in $\chi^{\mu_B}_2$. After this point, it directly goes to the non-interacting quark limit from above. As discussed earlier, both the parity doublet (red dashed line) and Hybrid QMN (black dash-dotted line) models feature the first-order liquid-gas phase transition, which, at $T=10$~MeV, is evident around $\mu_B = 917$~MeV. The parity doublet model develops a peak-like structure, indicating the chiral crossover, and then gradually goes to zero. The asymptotic behavior is due to the monotonically increasing repulsive interactions, through increasing $\omega$ mean-field expectation value. Hence, the parity doublet model does not saturate the non-interacting quark limit at high density. The Hybrid QMN model, on the other hand, shows the chiral peak at much lower value of the baryon chemical potential, resulting in a stronger transition. This is traced back to the momentum cutoff introduced for the nucleons through Eq.~(\ref{eq:fermi_dist_nucleons}). When the cutoff yields a non-negligible suppression of the Fermi-Dirac distribution, the Hybrid QMN result starts to visibly deviate from the parity doublet counterpart. Eventually, at higher values of the baryon chemical potential, the Hybrid QMN model result features a discontinuity connected with the change of the degrees of freedom, from nucleons to quarks. Clearly, the discontinuity is due to the jump in the expectation value of the bag field, which abruptly suppresses the nucleon density and enhances the corresponding density of the quarks, allowing for the proper asymptotic behavior to be achieved.

		The left panel of Fig.~\ref{fig:x2} shows the corresponding results on the $\chi_2$ observable for the case of $\alpha b_0=300$~MeV. In this case, all three models yield the first-order chiral phase transition. Similarly to the previous example, the parity doublet result does not saturate the proper asymptotic behavior, while the NJL result does. The second-order cumulant in the Hybrid QMN model deviates from the result obtained in the parity doublet model much earlier, immediately after the liquid-gas phase transition. Note that such behavior is expected, since the value of the $\alpha$~parameter in this case is the lowest that does not spoil the nuclear groundstate properties at zero temperature. In general, lower value of $\alpha$ yields stronger suppression of the thermal fluctuations of the nucleon degrees of freedom, which becomes relevant already at much lower values of the baryon chemical potential. As a result, it triggers the chiral phase transition at much lower $\mu_B$. At the same time, it allows for an earlier onset of the quark degrees of freedom, and eventually the deconfinement transition occurs at lower $\mu_B$ as well.

		By contrast, the second-order cumulant is not sensitive to the first-order deconfinement transition in the Hybrid QMN model, as seen in Fig~\ref{fig:x2}. This is connected with the fact that the deconfinement transition is driven solely by the expectation value of the $b$ field, which is generated through the bag potential defined in Eq.~(\ref{eq:bag_potential}). Since the underlying symmetry of the potential $V_b$ is discrete, the deconfinement is dominated by the {\it massive} scalar field $b$. To see this, let us consider $T=0$ limit for simplicity, and assume that the transition from nucleons to quarks happens at some high value of the baryon chemical potential. In this limit, we expect that the mean fields $\sigma$, $\omega$ $\rightarrow 0$. Hence, the energies of the nucleons and quarks can be ignored. In such limit, one obtains the approximated expression for the gap equation of the $b$ field from Eq.~(\ref{eq:gap_bag}), namely
		\begin{equation}\label{eq:approximated_gap_eq}
			\frac{\partial \Omega}{\partial b} = - \kappa_b^2b + \lambda_b b^3 + \frac{2}{\pi^2} b^2 \left(1 - 2\alpha^3\right) \mu_B = 0 \textrm.
		\end{equation}
		The above sets an upper limit for the $\alpha$ parameter, $\alpha_{\rm max} = 2^{-1/3}$. This implies $\alpha b_0 \lesssim 452$~MeV at $T=0$, which is consistent with the constraint introduced in Section~\ref{sec:parameters}.

		Moreover, Eq.~(\ref{eq:approximated_gap_eq}) can be solved for $b$. Taking into account only the positive solution, one gets the approximate expression,
		\begin{equation}\label{eq:approximate_b}
			b = \frac{-\left(1-2\alpha^3\right)\mu_B + \sqrt{\left(1-2\alpha^3\right)^2\mu^2_B + \pi^2\kappa_b^2\lambda_b}}{\pi^2\lambda_b} \textrm,
		\end{equation}
		and further, $\partial b/\partial \mu_B$, namely
		\begin{equation}\label{eq:approximate_derivative_b}
			\frac{\partial b}{\partial \mu_B} = \frac{1-2\alpha^3}{\pi^2\lambda_b}\left[ \frac{(1-2\alpha^3)\mu_B}{\sqrt{(1-2\alpha^3)^2\mu_B^2 + \pi^4\kappa_b^2\lambda_b}} - 1 \right] \textrm.
		\end{equation}
		It is evident that Eqs.~(\ref{eq:approximate_b}) and~(\ref{eq:approximate_derivative_b}) are smooth functions that do not exhibit any non-monotonic behavior and asymptotically, in the limit of high $\mu_B$, go to zero.

		Under the above approximation, the derivative of the gap equation (\ref{eq:approximated_gap_eq}) with respect to the bag field is
		\begin{equation}
			\frac{\partial^2 \Omega}{\partial b^2} = - \kappa_b^2 + 3\lambda_b b^2 + \frac{4}{\pi^2} b \left(1 - 2\alpha^3\right) \mu_B\textrm.
		\end{equation}
		From the above it is clear that the bag field is massive, and that there is indeed no additional soft mode to couple to the baryon number density other than the $\sigma$ mode. Hence, the higher-order cumulants do not develop any non-monotonic behaviors at the deconfinement transition.

\begin{figure}[t]
		\begin{center}
			\includegraphics{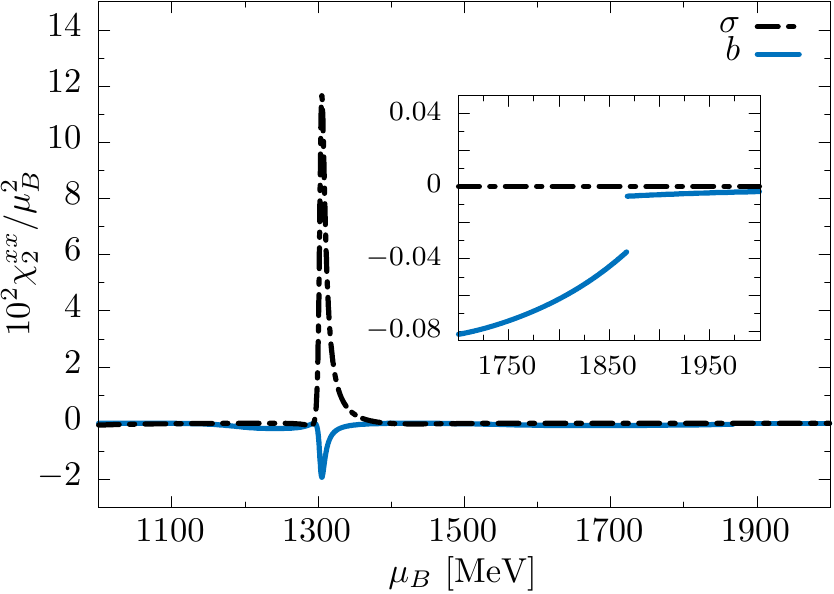}
			\caption{(Color online) The diagonal contributions to the second-order cumulant from the $\sigma$ and $b$ mean fields, for the case of $\alpha b_0=390$~MeV. The inset figure shows the details in the vicinity of the deconfinement transition.}
			\label{fig:x2_sigma_bag}
		\end{center}
		\end{figure}

		More details about the characteristics of the \mbox{second-order} cumulant can be illustrated by considering the individual contributions from the apparent $\mu_B$-dependence in the Fermi gas part and intrinsic one in the mean-field sector. We rewrite the cumulants in Eq.~(\ref{eq:cumulants_def}) as follows
		\begin{equation}\label{eq:x2_decomposition}
			\chi_2 = \sum_{x,y} \chi^{xy}_2= - \sum_{x,y} \frac{\partial^2 \Omega}{\partial x \partial y} \frac{\partial x}{\partial \mu_B} \frac{\partial y}{\partial \mu_B}\textrm,
		\end{equation}
		where $x,y=\mu_B$, $\sigma$, $\omega$, $b$. To see the explicit dependence and sensitivity of the \mbox{second-order} cumulant to the mean fields, we compare the diagonal terms from Eq.~(\ref{eq:x2_decomposition}). The contribution of the $\sigma$ and $b$ mean fields for the case of $\alpha b_0 = 390$~MeV are shown in Fig.~\ref{fig:x2_sigma_bag} normalized by $\mu_B^2$.
		
		The diagonal term $\chi_2^{\sigma\sigma}$ yields a positive contribution around the chiral phase transition, while it remains insensitive to the deconfinement transition (the inset plot of Fig.~\ref{fig:x2_sigma_bag}). This is because the sigma field does not play any role in activating the quarks. On the other hand, the term $\chi_2^{bb}$, shows a negative contribution and much weaker sensitivity to the chiral phase transition, and develops a jump due to the first-order deconfinement transition.
		
		We note that when the model is naively extended to higher-temperature domain, a critical point for the deconfinement transitions builds up, and the corresponding second-order cumulant exhibits an additional peak structure as well. This is because the $b$ field becomes a soft-mode at the second-order transition. Such extension is, however, beyond the current scope of the HQMN model, as it lacks, e.g., the implementation of the thermal fluctuations of mesons and gluons, which are relevant degrees of freedom at high temperature.
		
		\begin{figure*}[t]
		\begin{center}
			\includegraphics{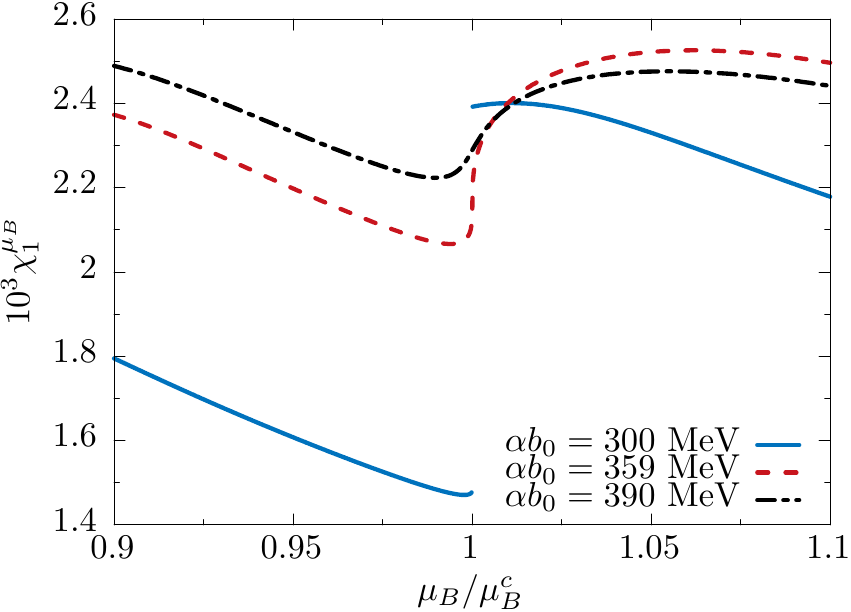}
			\includegraphics{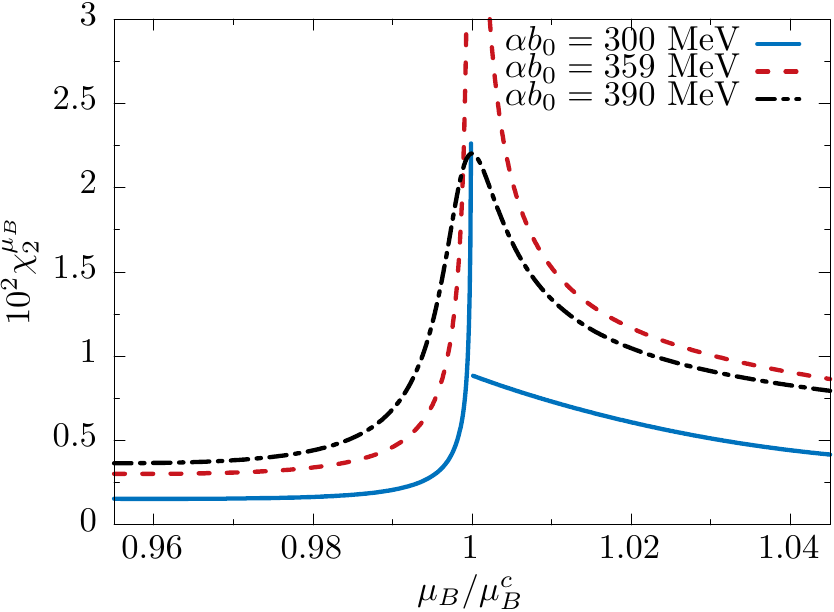}\\
			\includegraphics{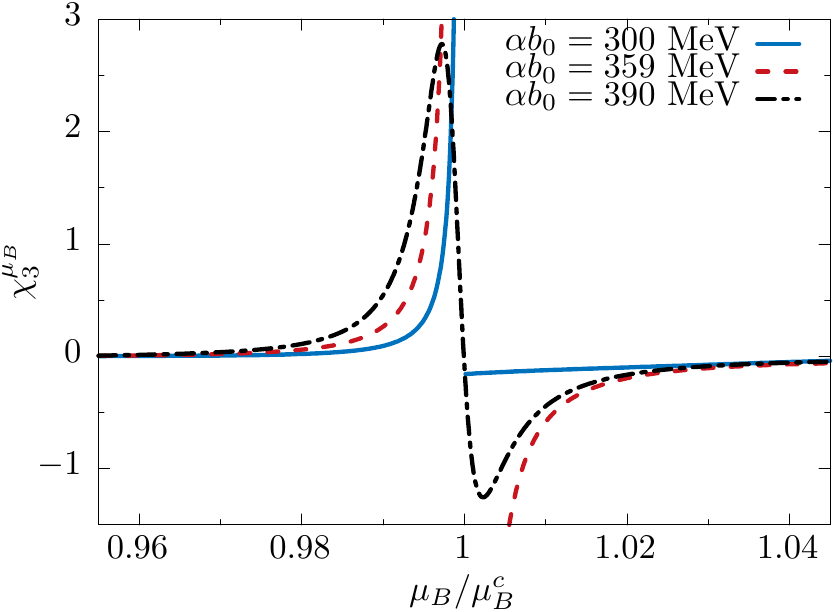}
			\includegraphics{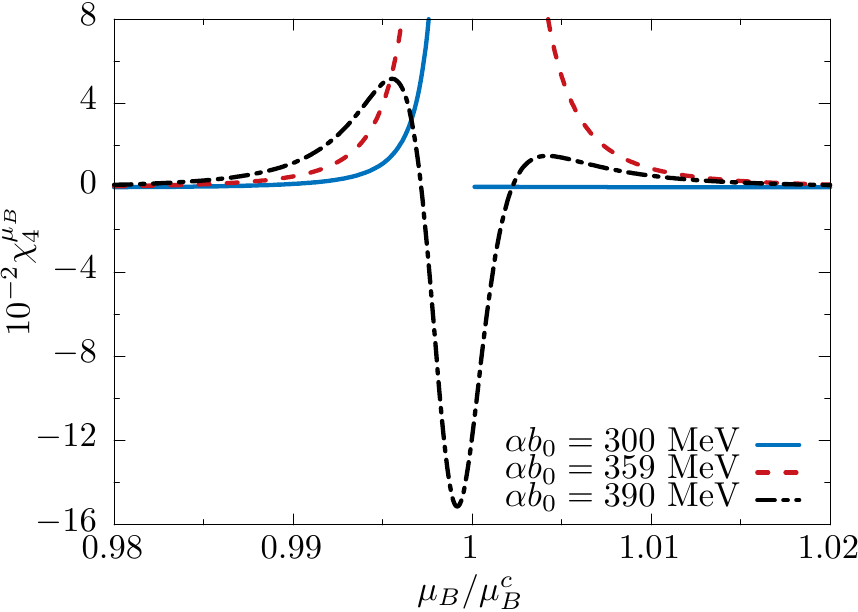}
			\caption{(Color online) First four cumulants in the Hybrid QMN model for different values of the parameter $\alpha$ plotted as a function of baryon chemical potential normalized by corresponding values of the chiral transition $\mu_B^c$ in each case.}
			\label{fig:x1234_all}
		\end{center}
		\end{figure*}
		
%%%%%%%%%%%%%%%%%%%%%%%%%%%%%%%%%%%%%%%%%%%%%%%%%%%%%%%%%%%%%%%
	\subsection{Higher-order cumulants}
%%%%%%%%%%%%%%%%%%%%%%%%%%%%%%%%%%%%%%%%%%%%%%%%%%%%%%%%%%%%%%%
	
		%{\bf [similar trend to chi2, comparison to PQM, criticality]}
		
		In Fig.~\ref{fig:x1234_all}, we show the first four cumulants of the net-baryon number density in the Hybrid QMN model, normalized by the baryon chemical potential in the vicinity of the chiral phase transition. Shown are the cumulants for different values of the $\alpha$ parameter, namely for the first-order transition (blue solid line), the second-order transition at the CP (red dashed line), and the crossover (black dash-dotted line). The \mbox{x-axis} is normalized by the corresponding values of the chiral transition point, $\mu_B^c$, in each scenario.
		
		In general, the first two cumulants, $\chi_1$ and $\chi_2$, stay positive at all values of the baryon chemical potential. This changes for higher orders, and the third- and fourth-order cumulants can both turn negative.
		
		%The cumulants $\chi_1^{\mu_B}$ and $\chi_2^{\mu_B}$ stay positive for all values of baryon chemical potential. This changes for higher orders and, near the chiral critical region, the third- and fourth-order cumulants can both turn negative.
		
		%third order

		%The third-order cumulant changes sign form positive in the chirally broken phase, to negative in the chirally restored phase. In all cases the cumulants rapidly increase in the chirally broken phase. For the first-order transition, it jumps to small negative values, while it diverges at both sides of the transition point. For the crossover, the cumulant increases in before the transition point and abruptly drops, exhibiting a dip at negative values.
		
		In the case of the first-order transition, the third-order cumulant, seen in the bottom left panel of Fig.~\ref{fig:x1234_all}, is discontinues and changes its sign from positive to small negative values at the chiral transition point. For the \mbox{second-order} transition, the cumulant diverges from both sides of the transition point and stays positive in the chirally broken phase, while it turns negative in the restored phase. In the case of the crossover transition, $\chi_3^{\mu_B}$ is continuous and rapidly increases in the vicinity of the transition, then abruptly drops, exhibiting a dip at negative values.
		
		%fourth order
		As seen in the bottom right panel of Fig.~\ref{fig:x1234_all}, the fourth-order cumulant exhibits similar characteristic structure to the third-order cumulant, while it stays positive for the first- and second-order case. For the crossover case, similarly to the third-order cumulant, in the chirally broken phase $\chi_4^{\mu_B}$ develops a peak, then suddenly drops and exhibit a dip at negative values. It then turns positive again, in contrast to $\chi_3^{\mu_B}$, and forms the second peak in the chirally restored phase. Note that, due to the deconfinement transition at higher values of baryon chemical potential, the cumulants reproduce the appropriate Stefan-Boltzmann limit. The signal of the transition is, however, less pronounced in the higher-order cumulants. This insignificance was already evident in the second-order cumulant.
		
		We note that the above characteristics, as expected, are qualitatively similar to the results obtained in different chiral models that incorporate the concept of statistical confinement in a different scheme, e.g., the Polyakov loop-extended quark-meson model~\cite{Skokov2}. This is because the transitions remain well separated at low temperatures, and so are the critical behaviors in the vicinity of each critical point. When the model is appropriately extended to higher temperatures, maximally three critical points could appear on the phase diagram. If they still remain separated, the cumulants, as well as their ratios, would exhibit a structure similar to the one discussed in the current work. If, on the other hand, the critical points would appear in a more complex way, e.g., closer to each other, a non-trivial structure could appear. We leave this is issue for a future work.
		
		%We note that the critical properties of a system with $O(4)$ and $Z(2)$ scaling have to be captured in effective models that implement chiral physics, irrespectively of the degrees of freedom.

%%%%%%%%%%%%%%%%%%%%%%%%%%%%%%%%%%%%%%%%%%%%%%%%%%%%%%%%%%%%%%%
\section{Summary and Conclusions}
\label{sec:conclusions}
%%%%%%%%%%%%%%%%%%%%%%%%%%%%%%%%%%%%%%%%%%%%%%%%%%%%%%%%%%%%%%%
	
	We have explored, under the mean-field-approximation, thermodynamics of the Hybrid Quark-Meson-Nucleon (Hybrid QMN) model at low temperature and various baryon chemical potential. The model describes nuclear matter in terms of the nucleon parity-doublers and mesons, and quark matter in the standard chiral-quark framework. The quark and nucleon sectors are connected via an auxiliary non-dynamical scalar field $b$, generated through a scalar potential $V_b$. The role of this field is to suppress unphysical thermal fluctuations of quarks at low $T$ and $\mu_B$ and, simultaneously, those of nucleons at high $T$ and $\mu_B$. This is achieved by the modified \mbox{Fermi-Dirac} distribution functions, where the $\alpha$~parameter is introduced to manipulate a size of the momentum cutoff. We emphasize that the inclusion of the cutoff does not spoil the universality and the critical behaviors.
	
	%The model was the fixed to reproduce the nuclear groundstate properties at $T=0$, We have used a high value of the chirally invariant mass $m_0=790$~MeV.
	
	We have studied the impact of the $\alpha$~parameter on the phase structure in the Hybrid QMN model at finite temperature. We find that the chiral and deconfinement transitions are always separated for a large $m_0$. The $\alpha$~parameter plays a crucial role in tuning the order of the chiral phase transition. Namely, the system undergoes a first-order chiral phase transition for low values of $\alpha$ and goes through a critical point, to eventually turn into a smooth crossover at higher $\alpha$. The deconfinement transition is, on the other hand, always of first order. This is due to the specific choice of the potential $V_b$, which by construction exhibits a first-order transition at low temperature. The separation of the two transitions might indicate the quarkyonic phase, where the quarks are confined, but the chiral symmetry is restored~\cite{Hidaka:2008yy, McLerran:2008ua, Andronic:2009gj}.
	
	%By performing numerical calculations of the critical exponent, we have confirmed the appropriate critical behavior of the second-order cumulant of the net-baryon number density at the critical point, which should be naturally encoded in chiral models.
	
	%We have considered a high value of the chirally invariant mass $m_0=790$~MeV, for which we have shown that, at low temperatures, the Hybrid QMN model phase structure is such that the chiral and deconfinement transitions are always separated. The order of the former transition is tuned with the $\alpha$ parameter.
	
	%Low values of the parameter yield a first-order transition, which goes trough a critial point and becomes smooth crossover for higher values of $\alpha$. The deconfinement transition on the other hand is always of first order
	
	%It yields a first-order phase transition for low values, which goes through a critical point, eventually turning into a smooth crossover at high values of the $\alpha$ parameter. The latter transition is on the other hand always of first order, which is driven by the expectation value of the $b$ field.
	
	We have also studied the critical behavior of the higher-order cumulants up to the fourth order, for different choice of the model parameter~$\alpha$. It is found that the impact of the suppression of the nucleon degrees of freedom on the chiral phase transition is twofold. First, it triggers the transition at much earlier baryon chemical potential, and second the transition is strengthened in comparison to the pure hadronic-model result. Since the changeover from nucleons to quarks is done self-consistently in the Hybrid QMN model, the observables properly saturate the asymptotic Stefan-Boltzmann limit. However, the deconfinement transition does not yield any non-monotonic behaviors.
	
	%strengthens and shifts the chiral phase transition towards lower values of the baryon chemical potential. The strengthening is particularly evident already in the second-order cumulant.
	
	%We have also discussed the asymptotic behavior of the first four cumulants, as well as their critical behavior in the vicinity of the chiral and deconfinement transitions.
	%We have indicated that, comparing to the purely hadronic parity doublet model, the suppression of the nucleon degrees of freedom streghtens and shifts the chiral phase transition towards lower values of the baryon chemical potential.
	
	In this study, we considered the two-flavored system, and it is a natural extension of the work to a theory including heavier flavors. The parity-doubling structure of baryons with three flavors was recently formulated in effective chiral approaches~\cite{Steinheimer:2011ea, Sasaki:2017glk}. It was shown that the masses of the baryon parity-doublers measured on lattice~\cite{Aarts:2015mma, Aarts:2017rrl} are modified to a large extent with the physical pion mass, in particular in the hyperon channels~\cite{Sasaki:2017glk}. It is of great interest to establish the equations of state of nuclear/hyperon matter in a physical setup.
	
	In view of the recent success of effective \mbox{quark-hadron} models in astrophysics~\cite{Benic:2014jia, Klahn:2015mfa, Kaltenborn:2017hus}, in particular in order to be consistent with current constraints, e.g., the observation of a massive two-solar-mass pulsar~\cite{Antoniadis:2013pzd}, repulsive interactions among quarks are essential. Moreover, in order to employ the Hybrid QMN model in astrophysical studies of compact stellar objects, such as neutron (or hybrid) stars, proto-neutron stars and supernovae~\cite{Pons:2001ar, Sagert:2008ka, Fischer:2010wp}, it is essential to extend the model to arbitrary isospin chemical potential. Further studies of these issues will be reported elsewhere.
	%However, these tasks extend beyond the scope of the present study, and we therefore leave them as a matter of future investigations.
	
	%In view of the recent success of the effective \mbox{quark-hadron} models~\cite{Benic:2014jia, Klahn:2015mfa, Kaltenborn:2017hus} in the astrophysical applications and fulfilling the constraints, such as the two-solar-mass neutron star~\cite{Antoniadis:2013pzd}, it would be interesting to examine the influence of a finite isospin chemical potential over the asymmetric dense matter in the Hybrid QMN model.
	
	%Repulsive interactions among quark degrees of freedom were so far not taken into account in the current approach. Hence, it would be instructive to study their influence on the chiral and the deconfinement transitions in the model. This could be essential for the stability of massive compact stars with quark cores and is currently under investigation.
	
	The potential of the $b$ field is so far not anchored to any QCD symmetry. It is indispensable to establish its dynamics in terms of a reliable symmetry. This will lead to a better understanding of its role as an order parameter for the deconfinement phase transition. In fact, it has been suggested that such a non-trivial order parameter of the center-flavor symmetry exists in a QCD-like theory compactified on a circle~\cite{Cherman:2017tey, Aitken:2017ayq}. This symmetry may be used to constrain further our model.
	
%%%%%%%%%%%%%%%%%%%%%%%%%%%%%%%%%%%%%%%%%%%%%%%%%%%%%%%%%%%%%%%
\acknowledgments
%%%%%%%%%%%%%%%%%%%%%%%%%%%%%%%%%%%%%%%%%%%%%%%%%%%%%%%%%%%%%%%

	The authors acknowledge fruitful discussions with P.~M.~Lo and K. ~Redlich. MM also thanks N.-U.~F.~Bastian, S.~Beni\'c, T.~Fischer, J.~Jankowski, and M.~Szyma\'nski for their helpful comments. This work is supported by the Polish National Science Center (NCN), under Maestro grant DEC-2013/10/A/ST2/00106. MM is also partly supported by the IFT grant no.~0420-2019-16.

%%%%%%%%%%%%%%%%%%%%%%%%%%%%%%%%%%%%%%%%%%%%%%%%%%%%%%%%%%%%%%%
\appendix
\section{Asymptotic behavior of the cumulants}
\label{sec:appendix}
%%%%%%%%%%%%%%%%%%%%%%%%%%%%%%%%%%%%%%%%%%%%%%%%%%%%%%%%%%%%%%%

	In the following, we discuss the asymptotic behavior of the observables constructed from the net-baryon number cumulants and their ratios, depending on the normalization.

	At low temperature and low baryon chemical potential the main degrees of freedom, due to confinement, are baryons, with baryon number $B=\pm 1$, while at high temperature quarks become dominant, with $B=\pm \frac{1}{3}$. When modeling the thermodynamics of QCD, especially when including the quark degrees of freedom, it is required that, at high temperature and high baryon chemical potential, the thermodynamic quantities reach the appropriate non-interacting quark gas limit. The general expression for the thermodynamic potential of the non-interacting gas of massless particles, carrying baryon quantum number $B$, reads
	\begin{equation}\label{eq:thermo_potential_noninteracting_quarks}
		\Omega = \frac{\gamma}{12 \pi^2} \left( B^4\mu_B^4 + 2\pi^2B^2\mu_B^2 T^2 + \frac{7}{15}\pi^4T^4\right) \textrm,
	\end{equation}
	with $\gamma$ being the spin degeneracy factor. The first four cumulants read
	\begin{subequations}\label{eq:free_cumulants}
	\begin{align}
		\chi_1 &= \frac{\gamma}{3\pi^2} \left( B^4\mu_B^3 + \pi^2 B^2 \mu_BT^2 \right) \textrm,\\
		\chi_2 &= \frac{\gamma}{3\pi^2} \left( 3B^4\mu_B^2 +\pi^2 B^2 T^2 \right) \textrm,\\
		\chi_3 &= \frac{2\gamma}{\pi^2}B^4\mu_B \textrm,\\
		\chi_4 &= \frac{2\gamma}{\pi^2}B^4 \textrm.\label{eq:4th_cumulant_free}
	\end{align}
	\end{subequations}
	Note that the fourth-order cumulant is already dimensionless and independent of the temperature and the baryon chemical potential. Moreover, it is directly proportional to the baryon quantum number $B$. Hence, $\chi_4$ itself can be a useful probe of the state of matter, in both hot and dense limit.
	
	Using the temperature-normalization scheme, defined in Eq.~(\ref{eq:temperature_normalization}), the asymptotic behavior in the \mbox{high-temperature} limit is the following
	\begin{subequations}\label{eq:cumulatns_T_hot_limit}
	\begin{align}
		\lim_{T\rightarrow \infty} \chi^T_1 & = 0\textrm,\\
		\lim_{T\rightarrow \infty} \chi^T_2 & = \frac{\gamma}{3}B^2\textrm,\\
		\lim_{T\rightarrow \infty} \chi^T_3 & = 0\textrm,\\
		\lim_{T\rightarrow \infty} \chi^T_4 & = \frac{2\gamma}{\pi^2}B^4\textrm.
	\end{align}
	\end{subequations}
	All of the four cumulants have well defined limits, but only the even ones depend on the baryon quantum number $B$. The odd cumulants vanish in the high temperature limit, owing to the baryon-antibaryon symmetry. On the other hand, in the limit of high baryon chemical potential, the first three cumulants go to infinity, and only the fourth one is proportional to $B$, namely
	\begin{subequations}\label{eq:cumulatns_T_dense_limit}
	\begin{align}
		\lim_{\mu_B\rightarrow \infty} \chi^T_1 & = \infty\textrm,\\
		\lim_{\mu_B\rightarrow \infty} \chi^T_2 & = \infty\textrm,\\
		\lim_{\mu_B\rightarrow \infty} \chi^T_3 & = \infty\textrm,\\
		\lim_{\mu_B\rightarrow \infty} \chi^T_4 & = \frac{2\gamma}{\pi^2}B^4\textrm.
	\end{align}
	\end{subequations}
	This suggests that, in order to probe the content of the QCD matter at high density, only the fourth cumulant can be utilized.
	
	The high-temperature limits of $\chi_n^{\mu_B}$, defined in Eq.~(\ref{eq:chemical_potential_normalization}), read
	\begin{subequations}\label{eq:cumulatns_u_hot_limit}
	\begin{align}
		\lim_{T\rightarrow \infty} \chi^{\mu_B}_1 & = \infty\textrm,\\
		\lim_{T\rightarrow \infty} \chi^{\mu_B}_2 & = \infty\textrm,\\
		\lim_{T\rightarrow \infty} \chi^{\mu_B}_3 & = \frac{2\gamma}{\pi^2}B^4\textrm,\\
		\lim_{T\rightarrow \infty} \chi^{\mu_B}_4 & = \frac{2\gamma}{\pi^2}B^4\textrm,
	\end{align}
	\end{subequations}
	and consequently, the limits of high baryon chemical potential,
	\begin{subequations}\label{eq:cumulatns_u_dense_limit}
	\begin{align}
		\lim_{\mu_B\rightarrow \infty} \chi^{\mu_B}_1 & = \frac{\gamma}{3\pi^4}B^4\textrm,\\
		\lim_{\mu_B\rightarrow \infty} \chi^{\mu_B}_2 & = \frac{\gamma}{\pi^4}B^4\textrm,\\
		\lim_{\mu_B\rightarrow \infty} \chi^{\mu_B}_3 & = \frac{2\gamma}{\pi^2}B^4\textrm,\\
		\lim_{\mu_B\rightarrow \infty} \chi^{\mu_B}_4 & = \frac{2\gamma}{\pi^2}B^4\textrm.
	\end{align}
	\end{subequations}
	Here, the situation is quite opposite. Namely, the high-density limit is well defined and are strictly proportional to the baryon quantum number $B$ for all four cumulants. The hot limit, on the other hand, is well defined only for the third and the fourth cumulants.
	
	Very useful are also the observables formed from ratios of the net-baryon density cumulants. Here, we discuss an example of the ratio of the fourth-order to the \mbox{second-order} cumulant, the so-called kurtosis~\cite{lat1}. At low temperature, where the Boltzmann approximation can be applied, the grand canonical ensemble can be approximated by \mbox{$\Omega \simeq - F(T) \cosh\left(B \mu_B / T \right)$}. From this, the kurtosis reads $R_{4,2} = B^2 / T^2$. At high temperature, one can use Eqs.~(\ref{eq:free_cumulants}) to obtain $R_{4,2} = 6B^2 / (3\mu_B^2B^2 + \pi^2 T^2)$. Using the normalization by temperature, the Boltzmann approximation yields
	\begin{equation}
		R^T_{4,2} = B^2 \textrm,
	\end{equation}
	and the high-temperature limit,
	\begin{equation}
		R^T_{4,2} = \frac{6}{\pi^2}B^2 \textrm,
	\end{equation}
	where the factor $6/\pi^2$ is due to the quantum statistics. The ratio $R_{4,2}^T$ is therefore roughly proportional to the square of the baryon number of the main degrees of freedom. Hence, it may serve as a very good probe for the proper determination of the state of QCD matter in the regime of low baryon chemical potential.
	
	The same quantity, however, turns out to be less useful in the low-temperature and dense regime. In the high-baron-chemical-potential limit,
	\begin{equation}\label{eq:kurtosis_limits_T_u}
		R^T_{4, 2} \xrightarrow[]{\mu_B\rightarrow \infty} 0 \textrm,
	\end{equation}
	the ratio is independent of the baryon quantum number $B$, and goes strictly to zero, regardless of which degrees of freedom are dominant.
	
	Similar result is obtained with normalization by the baryon chemical potential, where one finds that both hot and dense limits yield results independent of the content of matter, namely
	\begin{equation}\label{eq:kurtosis_limits_u}
	\begin{split}
		R^{\mu_B}_{4, 2} & \xrightarrow[]{T \rightarrow \infty} 0 \textrm,\\
		R^{\mu_B}_{4, 2} & \xrightarrow[]{\mu_B\rightarrow \infty} 2 \textrm.
	\end{split}
	\end{equation}
	In fact, it can be easily checked that all the ratios, defined through Eq.~(\ref{eq:chemical_potential_normalization}), yield limits independent of the baryon quantum number $B$. Nevertheless, still useful are the cumulants themselves, which at high density should approach the non-interacting gas counterpart results. Note that these are directly proportional to the baryon quantum number $B$ and therefore may be used to probe the state of QCD matter at high densities.
	
	In summary, the higher order cumulants of the net-baryon number density and their ratios may be utilized as very useful probes of the state of QCD matter. However, the appropriate observables and their normalization scheme have to chosen depending on the \mbox{($T$, $\mu_B$)-regime} of interest.

%%%%%%%%%%%%%%%%%%%%%%%%%%%%%%%%%%%%%%%%%%%%%%%%%%%%%%%%%%%%%%%

\end{document}